\documentclass[11pt,letterpaper]{article}
\usepackage[T1]{fontenc}
\usepackage[utf8]{inputenc}
\usepackage{lmodern,microtype}
\usepackage[margin=1in]{geometry}
\usepackage{amsmath,amssymb,amsthm,mathtools,bm}
\usepackage{stmaryrd}
\usepackage{booktabs,array,enumitem}
\usepackage{xcolor,aliascnt}
\usepackage{colortbl}
\usepackage{tikz}
\usepackage{float}
\usepackage{caption}
\usepackage[numbers,square]{natbib}
\usepackage[pagebackref,colorlinks=true,linkcolor=blue!55!black,citecolor=blue!55!black,urlcolor=blue!55!black]{hyperref}
\renewcommand*{\backrefalt}[4]{%
  \ifcase #1\relax
  \or (Cited on page~#2.)%
  \else (Cited on pages~#2.)%
  \fi
}
\usepackage[nameinlink,noabbrev]{cleveref}
\setlist{nosep}
\newtheorem{theorem}{Theorem}[section]
\theoremstyle{plain}
\newaliascnt{lemma}{theorem}
\newtheorem{lemma}[lemma]{Lemma}
\aliascntresetthe{lemma}
\crefname{lemma}{Lemma}{Lemmas}
\Crefname{lemma}{Lemma}{Lemmas}
\theoremstyle{plain}
\newaliascnt{proposition}{theorem}

\aliascntresetthe{proposition}
\crefname{proposition}{Proposition}{Propositions}
\Crefname{proposition}{Proposition}{Propositions}
\theoremstyle{plain}
\newaliascnt{corollary}{theorem}
\newtheorem{corollary}[corollary]{Corollary}
\aliascntresetthe{corollary}
\crefname{corollary}{Corollary}{Corollaries}
\Crefname{corollary}{Corollary}{Corollaries}
\theoremstyle{plain}
\newaliascnt{definition}{theorem}
\newtheorem{definition}[definition]{Definition}
\aliascntresetthe{definition}
\crefname{definition}{Definition}{Definitions}
\Crefname{definition}{Definition}{Definitions}
\newtheoremstyle{italicremark}{.5\topsep}{.5\topsep}{\itshape}{}{\itshape}{.}{.5em}{}
\theoremstyle{italicremark}
\newaliascnt{remark}{theorem}
\newtheorem{remark}[remark]{Remark}
\aliascntresetthe{remark}
\crefname{remark}{Remark}{Remarks}
\Crefname{remark}{Remark}{Remarks}
\crefname{theorem}{Theorem}{Theorems}
\Crefname{theorem}{Theorem}{Theorems}
\crefname{section}{Section}{Sections}
\crefname{figure}{Figure}{Figures}
\Crefname{figure}{Figure}{Figures}
\crefname{table}{Table}{Tables}
\Crefname{table}{Table}{Tables}
\newcommand{\R}{\mathbb R}
\newcommand{\Q}{\mathbb Q}
\newcommand{\N}{\mathbb N}
\newcommand{\E}{\mathbb E}
\newcommand{\FIXP}{\mathrm{FIXP}}
\newcommand{\PPAD}{\mathrm{PPAD}}
\newcommand{\LFIXP}{\mathrm{Linear}\text{-}\mathrm{FIXP}}
\newcommand{\Sol}{\operatorname{Sol}}
\newcommand{\Fix}{\operatorname{Fix}}

\DeclarePairedDelimiter{\clip}{\llbracket}{\rrbracket}

\newcommand{\SL}{\mathrm{SL}}
\newcommand{\A}{\mathcal A}

\newcommand{\Gc}{\mathcal C}

\title{FIXP-Completeness in Second-Price Auctions}
\author{Nobutaka Shimizu\footnote{Institute of Science Tokyo}}
\date{}

\hypersetup{pdftitle={FIXP-Completeness in Second-Price Auctions}}

\begin{document}
\pagenumbering{roman}
\pagestyle{empty}
\maketitle
\thispagestyle{empty}
\vspace{-1.2em}

\begin{abstract}
Computing equilibria is a fundamental algorithmic problem in game theory and economics. Although general frameworks are available for proving membership in $\FIXP$, the class capturing exact fixed-point computation, establishing $\FIXP$-completeness often remains challenging.
We prove $\FIXP$-completeness of exact equilibrium computation in three second-price auction models:
\begin{itemize}
\item pacing with return-on-investment constraints, with multiplier cap $1+\delta$ for every fixed rational $\delta>0$ [Filos-Ratsikas, Hansen, Høgh, and Hollender (2024)];
\item budget-constrained generalized second-price sponsored shopping [Hansen, Li, and Nie (2026)];
\item budget-constrained throttling [Liu (2026)].
\end{itemize}
These results resolve open questions on the exact complexity of automated bidding.

Our technical ingredient is a $\FIXP$-complete generalized-circuit problem using only clipped affine gates and the reciprocal $x\mapsto1/(1+\tau x)$, for any fixed positive rational $\tau$. This gate basis provides a common framework for the three auction reductions: simulating these gates in each model establishes $\FIXP$-hardness without directly implementing multiplication or squaring.
\end{abstract}

\newpage
\begingroup
\setlength{\parskip}{0pt}
\tableofcontents
\endgroup
\newpage
\pagenumbering{arabic}
\pagestyle{plain}

\section{Introduction}\label{sec:intro}
How difficult is it to find a solution that is guaranteed to exist? Fixed-point theorems guarantee equilibria and stable outcomes in many models, but do not by themselves provide efficient algorithms for finding them. The class $\PPAD$~\cite{Pap94} is central to the complexity of approximate equilibrium computation~\cite{CDT09,FRHH24}. Exact computation poses an additional issue: solutions can be irrational even when the input is rational. The class $\FIXP$~\cite{EY10} captures exact fixed-point computation using algebraic circuits. A $\FIXP$-completeness result identifies a problem as having the full computational difficulty of this class. See \cref{subsec:related-work} for further background on fixed-point complexity and equilibrium computation in auctions.

General frameworks establish membership in $\FIXP$ and $\PPAD$ for broad families of equilibrium problems~\cite{FRHH23,FRHH24}, but proving hardness for these classes can remain difficult. For example, the Hylland--Zeckhauser assignment mechanism has a $\PPAD$-complete approximation problem~\cite{VY25,CCPY22}, while exact computation belongs to $\FIXP$ and its $\FIXP$-hardness remains open~\cite{VY25}. Neither hardness of approximation nor the existence of irrational solutions supplies the reduction needed to prove $\FIXP$-hardness.

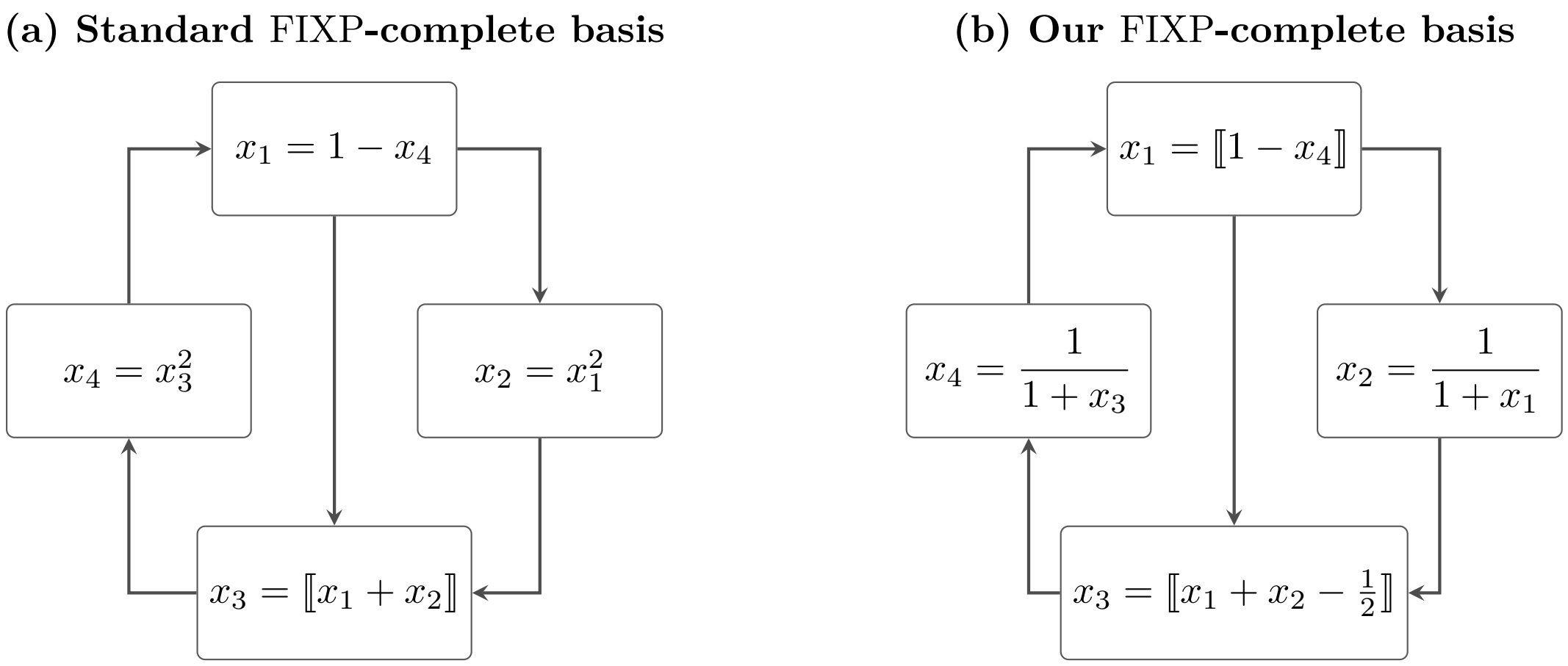
\begin{figure}[htbp]
\centering
\begin{tikzpicture}[
 font=\small,
 intro gate/.style={rectangle,rounded corners=2pt,draw=black!65,fill=white,minimum width=22mm,minimum height=12mm,inner sep=3pt,align=center},
 intro dependency/.style={->,draw=black!70,line width=.8pt,>=stealth}
]
% Both panels use the same diamond-shaped dependency graph.
\begin{scope}
\node[font=\small\bfseries] at (0,3.05) {(a) Standard $\FIXP$-complete basis};
\node[intro gate] (s1) at (0,2) {$x_1=1-x_4$};
\node[intro gate] (s2) at (1.85,0) {$x_2=x_1^2$};
\node[intro gate] (s3) at (0,-2) {$x_3=\clip{x_1+x_2}$};
\node[intro gate] (s4) at (-1.85,0) {$x_4=x_3^2$};
\draw[intro dependency] (s1.east) -| (s2.north);
\draw[intro dependency] (s2.south) |- (s3.east);
\draw[intro dependency] (s3.west) -| (s4.south);
\draw[intro dependency] (s4.north) |- (s1.west);
\draw[intro dependency] (s1.south) -- (s3.north);
\end{scope}
\begin{scope}[xshift=8.1cm]
\node[font=\small\bfseries] at (0,3.05) {(b) Our $\FIXP$-complete basis};
\node[intro gate] (r1) at (0,2) {$x_1=\clip{1-x_4}$};
\node[intro gate] (r2) at (1.85,0) {$\displaystyle x_2=\frac{1}{1+x_1}$};
\node[intro gate] (r3) at (0,-2) {$x_3=\clip{x_1+x_2-\tfrac12}$};
\node[intro gate] (r4) at (-1.85,0) {$\displaystyle x_4=\frac{1}{1+x_3}$};
\draw[intro dependency] (r1.east) -| (r2.north);
\draw[intro dependency] (r2.south) |- (r3.east);
\draw[intro dependency] (r3.west) -| (r4.south);
\draw[intro dependency] (r4.north) |- (r1.west);
\draw[intro dependency] (r1.south) -- (r3.north);
\end{scope}
\end{tikzpicture}
\caption{Two generalized circuits with the same dependency graph, using (a) a standard $\FIXP$-complete basis~\cite[Proposition~5.6]{FGHLP23} and (b) clipped affine and reciprocal gates from \eqref{eq:intro-gates}. An arrow $x_j\to x_i$ means that gate $i$ uses $x_j$ as an input; all variables lie in $[0,1]$.}
\label{fig:intro-reciprocal-circuit}
\end{figure}

A standard starting point for proving hardness is a \emph{generalized-circuit} problem~\cite{CDT09,Rub18,BHH21,FGHLP23}.
A generalized circuit is a system of gate equations whose input dependencies may contain directed cycles, as illustrated in \cref{fig:intro-reciprocal-circuit}. Each variable is the output of exactly one gate and may be used as an input by multiple gates. Each gate applies an operation to its inputs, usually one or two, and the task is to assign values $x_i\in[0,1]$ satisfying all gate equations simultaneously.\footnote{The original definition of $\FIXP$ uses fixed points of continuous maps represented by acyclic algebraic circuits~\cite{EY10}; the exact generalized-circuit formulation is equivalent~\cite{BHH21,FGHLP23}.} To use a generalized-circuit problem in a hardness reduction, we construct an equilibrium instance in polynomial time and specify how to recover a circuit solution from every equilibrium. Known $\FIXP$-complete gate sets include quadratic functions such as multiplication $(u,v)\mapsto uv$, squaring $u\mapsto u^2$~\cite[Proposition~5.6]{FGHLP23}, and the shifted product $(u,v)\mapsto(u+1)(v+1)/4$~\cite[Lemma~5.7]{FGHLP23}. By contrast, piecewise-linear functions that consist of affine operations and $\min,\max$ characterize $\PPAD$ through $\LFIXP=\PPAD$~\cite{EY10}.

For approximate problems, a reduction can replace a nonlinear operation by a piecewise-linear approximation if the resulting error stays within the allowed error bound~\cite{DGP09,CDT09}. An exact reduction must enforce the gate equation with zero error. Constructing equilibrium conditions that enforce multiplication or squaring can therefore require a separate argument for each target model~\cite{EY10,FGHLP23}, making $\FIXP$-hardness difficult to establish.
This motivates the following question:
\begin{quote}
\itshape Can we prove $\FIXP$-completeness of equilibrium problems without directly implementing quadratic functions in each target model?
\end{quote}

We answer this question affirmatively using generalized circuits with \emph{reciprocal gates}.
Fix a rational $\tau>0$ and write $\clip{t}=\min\{1,\max\{0,t\}\}$. An instance specifies one equation for each of $N$ variables $x_1,\ldots,x_N\in[0,1]$, of either form
\begin{equation}\label{eq:intro-gates}
 x_i=\clip{a_i x_{j_i}+b_i x_{k_i}+c_i}
 \qquad\text{or}\qquad
 x_i=\frac{1}{1+\tau x_{j_i}},
\end{equation}
where the coefficients are binary-encoded rationals and the input indices belong to $[N]=\{1,\ldots,N\}$. The right-hand sides define a continuous self-map of $[0,1]^N$, so a solution exists by Brouwer's fixed-point theorem. \Cref{fig:intro-reciprocal-circuit} compares a four-gate instance over a standard $\FIXP$-complete gate set with a four-gate instance over our gate set.

\begin{theorem}[see also \cref{thm:universal}]\label{thm:intro-framework}
For every fixed rational $\tau>0$, finding an exact solution of a generalized circuit using only the gates in \eqref{eq:intro-gates} is $\FIXP$-complete.\footnote{Indeed, $\FIXP$-completeness already holds with two gate functions: $H(u,v)=\min\{1,2-u-v\}$ and $R_1(u)=1/(1+u)$; see \cref{rem:two-gate-basis}.}
\end{theorem}

We apply \cref{thm:intro-framework} to automated bidding, where software chooses bids or participation probabilities while satisfying advertisers' financial constraints. Computing equilibria in these markets is central to predicting bidding behavior and evaluating revenue and welfare. The complexity of this task also helps assess whether automated bidding procedures can reach stable outcomes efficiently.

In a \emph{second-price auction} for a single item, the highest bidder wins and pays the highest competing bid, or zero if there is no competing bidder. When a bidder seeks to maximize its value for the item minus its payment and faces no financial constraint, bidding its value is optimal regardless of the other bids~\cite{Vic61}. This removes the need to predict competitors' bids, unlike a \emph{first-price auction}, where the winner pays its own bid.

The models involve multiple items and financial constraints that apply across items. Spending on one item can therefore affect how much a bidder can spend on others. Thus bidding one's value for each item is not necessarily optimal under these constraints. This makes equilibrium computation more challenging.

We study three second-price auction models: pacing in separate second-price auctions with return-on-investment (ROI) constraints~\cite{LT24,FRHH24}, budget-constrained autobidding in generalized second-price (GSP) sponsored shopping~\cite{HLN26}, and budget-constrained throttling in second-price auctions~\cite{CKK21,Liu26}. Prior work explicitly poses exact $\FIXP$-completeness as an open question for each model~\cite{FRHH24,HLN26,Liu26}. Using our reciprocal-gate framework based on \cref{thm:intro-framework}, we resolve these questions by proving $\FIXP$-completeness for all three problems.

\subsection{Pacing Equilibria in ROI-constrained Auctions}\label{subsubsec:intro-roi}
In this model~\cite{LT24,FRHH24}, each bidder assigns a value to each advertising opportunity (\emph{item}) and seeks to maximize the total value it acquires. Its \emph{ROI constraint} requires this total value to be at least a specified positive multiple of its total payment.
We consider the case in which total payment is at most total acquired value.

\emph{Pacing} controls all of a bidder's bids through one scalar, its \emph{multiplier}, so it can adjust its bidding across items to meet its financial constraint. Here each bidder chooses its own multiplier between one and the cap $1+\delta$, for a fixed rational $\delta>0$, and multiplies its value for every item by that same factor to obtain its bids. Each item is allocated among its highest bidders at the second-highest bid, with tied bidders allowed to share the item and pay in proportion to their shares. At a \emph{second-price ROI pacing equilibrium}, each bidder's total payment is at most the value of its allocation, and any bidder whose payment is less than that value uses multiplier $1+\delta$. The task is to find the multipliers and allocation from the bidders' values.

Li and Tang~\cite{LT24} prove $\PPAD$-hardness of approximate equilibrium computation even for constant approximation parameters. For exact computation, Filos-Ratsikas, Hansen, H\o gh, and Hollender~\cite[Section~7]{FRHH24} prove $\FIXP$-membership and give rational instances in which every equilibrium has an irrational multiplier. They explicitly ask whether exact equilibrium computation is $\FIXP$-complete. The following theorem answers this question for cap $1+\delta$ with any fixed rational $\delta>0$.

\begin{theorem}[Informal]\label{thm:intro-roi}
For every fixed rational $\delta>0$, computing an exact second-price ROI pacing equilibrium with cap $1+\delta$ from the bidders' values is $\FIXP$-complete.
\end{theorem}
The formal model and proof appear in \cref{sec:roi} (\cref{def:roi-pacing,thm:roi-complete}).

\subsection{Autobidding Equilibria in Sponsored Shopping}\label{subsubsec:intro-gsp}
In \emph{sponsored shopping}~\cite{DLPST26,HLN26}, sellers bid to have their products displayed in advertising positions (\emph{slots}), for example alongside shopping search results. Each bidder is a seller and may own several products, called \emph{items}, so it can occupy several slots. An item's value is its seller's value for a click on its advertisement. The seller's \emph{budget} limits its expected total payment. An \emph{autobidder} chooses bids for a seller using a multiplier between zero and one: it multiplies the values of all the seller's items by that same factor.

Items are ranked in nonincreasing order of bids, and the highest-ranked items fill the available slots. The \emph{generalized second-price} (GSP) rule extends second-price pricing to ranked advertising positions~\cite{EOS07}. In the sponsored-shopping model, an item's price is the first lower-ranked bid belonging to a different seller, or zero if no such item exists; other items of the same seller are skipped. We consider one slot per item, with all items displayed and each item's payment equal to its price.\footnote{In general, each slot has a \emph{click-through rate}, which is the probability that a user clicks an item displayed there. Values and payments are considered in expectation over clicks, so an item's value and price are multiplied by its slot's click-through rate. Here, however, we prove $\FIXP$-completeness even in the simple case where all click-through rates are fixed to one.}

A \emph{budget-constrained autobidding equilibrium} consists of multipliers and a possibly randomized ordering of items with equal bids. Each bidder's expected total payment, averaging over this ordering, must be at most its budget, and any bidder with unused budget must use multiplier one. The task is to find multipliers for which ties can be resolved to satisfy these conditions, given item ownership, values, and bidder budgets.

Hansen, Li, and Nie~\cite{HLN26} prove $\FIXP$-membership for exact equilibrium computation and $\PPAD$-completeness for inverse-polynomial approximation, under both budget and ROI constraints. They also give rational instances in which every exact equilibrium has an irrational multiplier. They explicitly ask whether exact equilibrium computation is $\FIXP$-hard~\cite[Section~7]{HLN26}. The following theorem answers this question for the budget-constrained model.

\begin{theorem}[Informal]\label{thm:intro-gsp}
With one slot per item and all click-through rates equal to one, computing a multiplier vector of an exact budget-constrained autobidding equilibrium in GSP sponsored shopping from item ownership, values, and bidder budgets is $\FIXP$-complete.
\end{theorem}
The formal model and proof appear in \cref{sec:gsp} (\cref{def:abe,thm:gsp-complete}).

\subsection{Throttling in Second-Price Auctions}\label{subsec:intro-throttling}
In \emph{throttling}~\cite{CKK21}, each bidder has fixed bids for the items and a positive budget. Each bidder, or an algorithm acting on its behalf, chooses a participation probability between zero and one to control its expected spending. With this probability, the bidder participates in the auctions for all items; otherwise, it participates in none. Participation decisions are independent across bidders. Each item is sold by the second-price rule among participating bidders. At an \emph{exact second-price throttling equilibrium}, each bidder's expected total payment is at most its budget, and any bidder with unused budget participates with probability one. The task is to find these participation probabilities from the bids and budgets.

Chen, Kroer, and Kumar~\cite{CKK21} prove $\FIXP$-membership for exact equilibrium computation and $\PPAD$-completeness for approximation with a sufficiently small constant error parameter. They also give rational instances in which every exact equilibrium has an irrational participation probability. Liu~\cite{Liu26} extends the approximation-hardness result to every fixed error parameter in $(0,1)$ and explicitly asks whether exact equilibrium computation is $\FIXP$-complete~\cite[Section~6]{Liu26}. The following theorem answers this question.

\begin{theorem}[Informal]\label{thm:intro-throttling}
Computing an exact budget-constrained throttling equilibrium in second-price auctions is $\FIXP$-complete, even when each item has at most three positive bids and these bids are pairwise distinct.
\end{theorem}
The formal model and proof appear in \cref{sec:throttling} (\cref{def:throttling,thm:throttling-complete}).

\subsection{Background and Related Work}\label{subsec:related-work}

\paragraph{PPAD, FIXP, and Equilibrium Computation}\label{subsubsec:related-fixed-points}
Nash's theorem guarantees that every finite game has a mixed Nash equilibrium, but does not provide an efficient algorithm for finding one. This motivates the study of \emph{total search problems}, in which every valid input has a solution: the challenge is finding a solution, rather than deciding whether one exists. Papadimitriou~\cite{Pap94} introduced $\PPAD$ to study a class of discrete total search problems arising from constructive fixed-point arguments.

Chen, Deng, and Teng~\cite{CDT09} prove $\PPAD$-completeness of exact Nash equilibrium computation in rational two-player games. Approximation hardness has also been studied extensively, from exponentially small and inverse-polynomial additive error~\cite{DGP09,CDT09} to constant error under several game models and approximation notions~\cite{Das13,Rub18,DFHM24}.

Beyond Nash equilibria, $\PPAD$-completeness results span competitive markets~\cite{CPY17}, one-sided matching~\cite{CCPY22}, fair division and cake cutting~\cite{KRS26}, and first- and second-price auctions~\cite{FGHLP23,CKK24}. These results concern exact or approximate solutions, depending on the model.

Exact computation presents an additional issue: rational games with three or more players may have only irrational equilibria. To study the complexity of computing such solutions exactly, Etessami and Yannakakis~\cite{EY10} introduced $\FIXP$. Its defining task is to find $\bm{x}=F(\bm{x})$ for a continuous self-map of a compact rational polytope, represented by a circuit using rational constants and arithmetic operations; see \cref{def:fixp,def:sl} for the formal definition and reductions. Their central result is that computing an exact mixed Nash equilibrium of a three-player finite game is $\FIXP$-complete. They also prove that the piecewise-linear restriction of $\FIXP$ equals $\PPAD$, connecting exact nonlinear fixed-point computation to its piecewise-linear counterpart. For more details on the fixed-point foundations of $\PPAD$ and $\FIXP$, see the survey by Yannakakis~\cite{Yan09}.

%Two meanings of approximation must also be distinguished. An $\varepsilon$-Nash equilibrium permits each player to gain at most $\varepsilon$ by deviating; an $\varepsilon$-approximation to an exact equilibrium must differ by at most $\varepsilon$ in every coordinate from some exact equilibrium. The former need not satisfy the latter. Etessami and Yannakakis~\cite{EY10} formalize the second task through $\FIXP_a$, the discrete approximation counterpart of $\FIXP$. Likewise, approximation results for auction and matching equilibria concern the specific relaxed equilibrium conditions in the cited papers.

Beyond finite games, $\FIXP$-completeness results span competitive markets~\cite{GMVY17}, Bayesian first-price auctions~\cite{FGHLP23}, and envy-free cake cutting with general valuations and computational versions of topological existence theorems~\cite{FRHH23}. Further connections between fair division and topology include $\FIXP$-hardness for exact consensus halving and the Borsuk--Ulam problem~\cite{DFMS21}.

Important open $\FIXP$-completeness questions span matching, markets, and fair division. Examples include exact equilibrium computation for the Hylland--Zeckhauser mechanism for one-sided matching~\cite{HZ79,VY25}, exact equilibrium computation in Arrow--Debreu exchange markets with constant-elasticity-of-substitution (CES) utilities~\cite{CPY17}, and contiguous envy-free cake cutting with additive valuations represented by algebraic circuits~\cite{FRHH23}.\footnote{For rational piecewise-constant additive valuations, exact contiguous envy-free cake cutting is $\PPAD$-complete~\cite{KRS26}.} In each case, $\FIXP$-membership is known, while the cited works leave $\FIXP$-hardness open. Our reciprocal-gate framework does not resolve these questions.

The membership side has a more general methodology. Filos-Ratsikas, Hansen, H\o gh, and Hollender~\cite{FRHH23,FRHH24} show how to incorporate suitably represented convex optimization problems into fixed-point constructions, obtaining $\FIXP$-membership and, under additional conditions ensuring suitable rational solutions, $\PPAD$-membership.
Our contribution concerns the complementary hardness direction: a restricted nonlinear gate basis that can be implemented through auction equilibrium conditions.

\paragraph{First-Price Auctions}\label{subsubsec:related-first-price}
In a first-price auction, bidders trade off the probability of winning against the payment upon winning. With private information, a pure Bayes--Nash equilibrium specifies a bid as a function of each bidder's private value, maximizing expected utility given its beliefs and the other bidding functions. For further background on auction theory and equilibrium bidding, see Klemperer's survey~\cite{Kle99}. Filos-Ratsikas, Giannakopoulos, Hollender, Lazos, and Po\c{c}as~\cite{FGHLP23} prove $\PPAD$-completeness for approximate and $\FIXP$-completeness for exact equilibrium computation in a Bayesian first-price auction model. Further $\PPAD$-completeness results concern approximation with common priors and an input-specified tie-breaking rule~\cite{CP23}, and approximate mixed equilibria with discrete values and bids under subjective priors~\cite{FGHK26} or a correlated common prior~\cite{CCCPS26}.

Under budget constraints, first-price pacing has unique equilibrium multipliers, computable in polynomial time through an Eisenberg--Gale convex program~\cite{CKPSSSW22}. First-price throttling has a unique equilibrium and admits polynomial-time approximation, although the exact equilibrium may be irrational~\cite{CKK21}.

\paragraph{Second-Price Auctions and Autobidding}\label{subsubsec:related-second-price}
For budget-constrained second-price pacing, Conitzer, Kroer, Sodomka, and Stier-Moses~\cite{CKSS22} establish existence and show that selecting a welfare- or revenue-maximizing equilibrium is NP-hard. Chen, Kroer, and Kumar~\cite{CKK24} prove that finding any equilibrium is $\PPAD$-complete, for both exact computation and inverse-polynomial approximation. An exact equilibrium with rational coordinates of polynomial bit length always exists. Chen and Li~\cite{CL26} strengthen the hardness result to constant approximation, and Liu~\cite{Liu26} gives tight bounds. Polynomial-time algorithms are known when either the number of bidders~\cite{YWL26} or the number of goods~\cite{HYWL26} is constant.

For ROI-constrained autobidding in sponsored shopping, D\"utting, Li, Paes Leme, Spendlove, and Teng~\cite{DLPST26} establish existence and welfare guarantees. Our result in \cref{thm:intro-gsp} concerns budget constraints; the $\FIXP$-hardness of computing an exact ROI-constrained equilibrium remains open~\cite{HLN26}.

The use of multipliers is also supported by results on optimal bidding. In second-price auctions, an ROI-constrained bidder facing fixed competing bids can maximize its total acquired value using a multiplier, provided it may choose its share of each item on which it ties for the highest bid~\cite[Proposition~1]{LT24}. Related work studies optimal bidding and welfare under more general financial constraints~\cite{ABM19}, and algorithms that jointly handle budget and ROI constraints~\cite{BBFLMSW24}. For models with continuously distributed values, convergence guarantees for pacing and throttling have been obtained under smoothness and tail conditions~\cite{CMY26}. For a broader survey of autobidding algorithms, equilibria, and auction design, see~\cite{AutobiddingSurvey24}.

\section{Proof Overview}\label{sec:overview}

\subsection{The Challenge of Exact Gate Simulation}\label{subsec:overview-challenge}
Approximate gate constraints can already capture $\PPAD$-hardness using only piecewise-linear functions, meaning functions that are affine on each of finitely many regions. For example, for some fixed $\delta>0$, finding values in $[0,1]$ that satisfy every gate equation within additive error $\delta$ is $\PPAD$-complete even with only $u\mapsto1-u$ and $(u,v)\mapsto\min\{1,u+v\}$~\cite[Proposition~5.5]{FGHLP23}. Adding $u\mapsto u^2$ makes the exact version $\FIXP$-complete~\cite[Proposition~5.6]{FGHLP23}. This distinction also appears in the characterization $\LFIXP=\PPAD$ of Etessami and Yannakakis~\cite[Section~5]{EY10}: restricting $\FIXP$ circuits to rational affine operations and $\min,\max$ gives precisely $\PPAD$. Thus requiring zero error in a piecewise-linear circuit problem does not by itself establish $\FIXP$-hardness.

The second-price rule does not directly supply the missing nonlinear arithmetic. For $n\geq2$, the second-highest bid, counting tied bids separately, satisfies
\[
 \operatorname{2ndmax}(b_1,\ldots,b_n)
 =\max_{1\leq i<j\leq n}\min\{b_i,b_j\}.
\]
This is piecewise-linear, and any finite composition of this operation with affine operations remains piecewise-linear. It therefore cannot compute $u^2$ or $uv$ exactly on the whole input domain. A reduction must use the full equilibrium conditions to enforce such relations, rather than only compose the operations that determine prices.

A related obstacle is explicit in the first-price auction reduction of~\cite[Section~5.2]{FGHLP23}: it is ``unclear how to simulate a $G_+$-gate or a $G_\times$-gate,'' referring to clipped addition and multiplication. That reduction instead implements $\phi(u,v)=(u+1)(v+1)/4$ together with clipped doubling and $u\mapsto1-u$, obtaining a $\FIXP$-complete exact circuit problem.

\subsection{Reciprocal Gate}\label{subsec:intro-framework}
We start from the generalized-circuit problem with gates 
\[
u\mapsto 1-u, \quad (u,v)\mapsto\clip{u+v}, \quad\text{and}\quad u\mapsto u^2,
\]
which is known to be $\FIXP$-complete~\cite[Proposition~5.6]{FGHLP23}.
We call the first type of gate in~\eqref{eq:intro-gates} a \emph{clipped affine gate} and the second a \emph{reciprocal gate}.
Our goal is to implement the three gates above using only these two gate types.
The gates $u\mapsto1-u$ and $(u,v)\mapsto\clip{u+v}$ are already clipped affine gates, so it suffices to implement squaring.

A key observation is that, for $0<t<1$, it holds that
\begin{equation}\label{eq:overview-hua}
 t^2=t-\left(\frac1t+\frac1{1-t}\right)^{-1}.
\end{equation}
This is a special case of Hua's identity~\cite{Hua49,Mattarei05}.
In our context, \eqref{eq:overview-hua} expresses a square using affine operations and reciprocals, but is undefined at the endpoints and can produce intermediate values exceeding one. To stay in $[0,1]$, we use the gates from~\eqref{eq:intro-gates}, writing $R_{\tau}(u)=1/(1+\tau u)$ for the reciprocal gate with rational $\tau>0$.

For example, to compute $x\mapsto x^2$ for $x\in[0,1]$ while keeping all intermediate values in $[0,1]$, first form

\[
 r_+=R_1\left(\frac{1+x}{2}\right),\qquad r_-=R_1\left(\frac{1-x}{2}\right),\quad\text{and}\quad
 s=\frac{3}{4}(r_++r_-)-1.
\]
These values lie in $[0,1]$ and satisfy
\[
 s=\frac{x^2}{9-x^2}\in\left[0,\frac{1}{8}\right]\quad\text{and}\quad x^2=9(1-R_1(s)).
\]
\Cref{lem:square} extends this construction to every $\tau>0$. Replacing each squaring gate preserves the source solutions and proves \cref{thm:intro-framework}.

\subsection{Inverse Pseudogate}\label{subsec:overview-auctions}
We now explain how to obtain the reciprocal from a suitable nonlinear response.

The key idea is to first use an auction construction to enforce a gate $x=G(u)$ whose inverse has the form
\begin{equation}\label{eq:overview-inverse-form}
 G^{-1}(x)=a+bx+cR_{\tau}(x),
\end{equation}
where $a,b,c,\tau$ are fixed rationals with $c\ne0$ and $\tau>0$. 
The first step is to implement $G^{-1}$ using the following inverse-gate construction.
\begin{lemma}[Informal]\label{lem:overview-inverse}
Let $G:[0,1]\to[0,1]$ be any continuous strictly decreasing bijection. Using auxiliary variables and directed cycles, gates for $G$ and clipped affine operations can implement the inverse gate $x\mapsto G^{-1}(x)$ exactly.
\end{lemma}
See \cref{lem:inverse} for the formal statement. On input $x\in[0,1]$, the construction uses auxiliary variables $y,z\in[0,1]$ satisfying
\[
 y=G(z)\quad\text{and}\quad z=\clip{z+y-x}.
\]
Together, the two equations force $y=x$. For $0<z<1$, the clipping equation gives $z=z+y-x$; at $z=0$ or $z=1$, the equation $y=G(z)$ and the endpoint values $G(0)=1$, $G(1)=0$ give the same conclusion. Hence each input $x$ has the unique extension $y=x$, $z=G^{-1}(x)$. This is the pseudogate principle of~\cite[Section~3.2]{FRHH23}: the auxiliary fixed-point equations enforce the required function at every solution.

Once $z=G^{-1}(x)$ is available, a clipped affine gate implements the reciprocal gate by outputting $\clip{(z-a-bx)/c}=R_{\tau}(x)$. Clipping does not change the value because $R_{\tau}(x)\in[0,1]$.

This construction replaces each reciprocal gate in the circuits of \cref{thm:intro-framework} by gates for $G$ and clipped affine gates: every original solution extends to the new circuit, and every new solution projects to an original solution. Finding an exact solution to a generalized circuit using only clipped affine gates and $G$ is $\FIXP$-complete, as formalized in \cref{cor:extraction}.

The following lemma connects this framework to the three auction models. ROI pacing and GSP sponsored shopping use the inverse pseudogate, whereas second-price throttling realizes the reciprocal directly. Recall that $R_1(u)=1/(1+u)$ is the reciprocal gate in \eqref{eq:intro-gates} with $\tau=1$.

\begin{lemma}\label{lem:overview-auction-simulation}
There exist continuous strictly decreasing bijections $G_{\mathrm{ROI}},G_{\mathrm{GSP}}:[0,1]\to[0,1]$, each with an inverse of the form \eqref{eq:overview-inverse-form}.
\begin{itemize}[itemsep=3pt]
 \item Every generalized circuit using clipped affine gates and $G_{\mathrm{ROI}}$ can be transformed in polynomial time into an ROI pacing instance.
 \item Every generalized circuit using clipped affine gates and $G_{\mathrm{GSP}}$ can be transformed in polynomial time into a budget-constrained GSP sponsored-shopping instance.
 \item Every generalized circuit using clipped affine gates and $R_1$ can be transformed in polynomial time into a second-price throttling instance.
\end{itemize}
In each case, every equilibrium of the constructed instance yields a solution of the original circuit.
\end{lemma}
For the auction constructions, \cref{lem:auction-affine} first replaces clipped affine gates by the constant $1$ and $H(u,v)=\min\{1,2-u-v\}$. The constructions and their correctness proofs appear in \cref{sec:roi,sec:gsp,sec:throttling}.

\subsection{A Reciprocal Relation in Throttling}\label{subsec:overview-throttling}
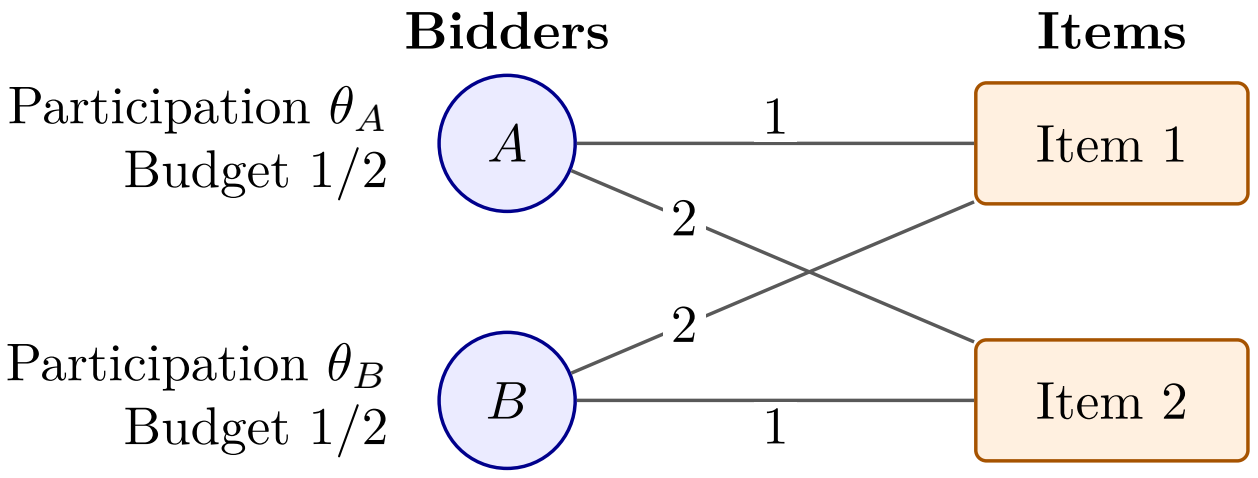
\begin{figure}[htbp]
\centering
\begin{tikzpicture}[
 font=\small,
 bidder/.style={circle,draw=blue!55!black,fill=blue!8,minimum size=9mm,inner sep=2pt,line width=.65pt},
 item/.style={rectangle,rounded corners=2pt,draw=orange!65!black,fill=orange!12,minimum width=18mm,minimum height=8mm,inner sep=2pt,line width=.65pt},
 bid/.style={draw=black!65,line width=.65pt},
 amount/.style={fill=white,inner sep=1.5pt}
]
\node[font=\small\bfseries] at (0,1.6) {Bidders};
\node[font=\small\bfseries] at (4,1.6) {Items};
\node[bidder] (a) at (0,.85) {$A$};
\node[bidder] (b) at (0,-.85) {$B$};
\node[item] (first) at (4,.85) {Item 1};
\node[item] (second) at (4,-.85) {Item 2};
\node[anchor=east,align=right] at (-.65,.85) {Participation $\theta_A$\\Budget $1/2$};
\node[anchor=east,align=right] at (-.65,-.85) {Participation $\theta_B$\\Budget $1/2$};
\draw[bid] (a) -- node[amount,above] {$1$} (first);
\draw[bid] (a) -- node[amount,pos=.28] {$2$} (second);
\draw[bid] (b) -- node[amount,pos=.28] {$2$} (first);
\draw[bid] (b) -- node[amount,below] {$1$} (second);
\end{tikzpicture}
\caption{The two-bidder throttling example. Circles denote bidders, rectangles denote items, and edge labels are fixed bids.}
\label{fig:throttling-reciprocal-example}
\end{figure}

The following example suggests why our reciprocal framework is well suited to second-price auctions. In the second-price throttling model of \cref{subsec:intro-throttling}, we consider the reciprocal gadget used by Chen, Kroer, and Kumar~\cite[Appendix~B.3]{CKK21} in their NP-hardness proof for revenue maximization. There are two bidders $A,B$ and two items, as shown in \cref{fig:throttling-reciprocal-example}. Their bid vectors, in item order, are $(1,2)$ and $(2,1)$, respectively, and each bidder has budget $1/2$. Bidder $U\in\{A,B\}$ participates with probability $\theta_U\in[0,1]$, independently of the other bidder. Each item is awarded to the participating bidder with the highest bid, who pays the other bidder's bid (second-price) if both participate and zero otherwise.

By independence, both bidders participate with probability $\theta_A\theta_B$. When they do, $A$ wins the second item and $B$ wins the first, each paying $1$; otherwise, both pay zero. Thus the expected total payments of $A$ and $B$ are
\[
 \E[\text{total payment of }A]
 =\E[\text{total payment of }B]
 =1\cdot\theta_A\theta_B=\theta_A\theta_B.
\]
A pair $(\theta_A,\theta_B)\in[0,1]^2$ is an exact throttling equilibrium (see \cref{def:throttling} for the formal definition) precisely when
\[
 \begin{aligned}
  &\theta_A\theta_B\leq\frac12\quad\text{and}\\
  &\theta_A\theta_B<\frac12
  \quad\Longrightarrow\quad \theta_A=\theta_B=1.
 \end{aligned}
\]
If $\theta_A\theta_B<1/2$, the last condition gives $\theta_A\theta_B=1$, contradicting $\theta_A\theta_B\leq1/2$. Hence every exact equilibrium satisfies
\[
 \theta_A\theta_B=\frac12.
\]
This simulates the reciprocal function $u\mapsto 1/(2u)$ on $[1/2,1]$.
To simulate an entire circuit, we must also account for the payments that arise when bidders are used by other gates; the construction in \cref{sec:throttling} handles these additional payments.

\section{Preliminaries}\label{sec:prelim}
For $N\in\N$, write $[N]=\{1,\ldots,N\}$. We use bold symbols for vectors and matrices, and ordinary symbols for their scalar entries, as in $\bm{x}=(x_i)_{i\in[N]}$. For $t\in\R$, let $[t]_+:=\max\{t,0\}$ denote its positive part. Recall that $\clip{t}=\min\{1,[t]_+\}$ denotes clipping to $[0,1]$. We write $\Sol_\Pi(I)$ for the solution set of an instance $I$ of $\Pi$ and $\Fix(F)=\{\bm{x}:F(\bm{x})=\bm{x}\}$ for the set of fixed points of a function $F$.

Rational inputs use binary numerators and denominators; input size is the total bit length of the instance description.

\begin{definition}[$\FIXP$]\label{def:fixp}
A \emph{basic $\FIXP$ instance} specifies a nonempty compact \emph{rational polytope}\footnote{A rational polytope is a bounded set defined by finitely many linear inequalities with rational coefficients.} $D$ and an algebraic circuit for a continuous map $F:D\to D$. An \emph{algebraic circuit} is a finite directed acyclic circuit whose gates apply the operations listed below. The circuit uses rational constants and $+,-,\times,\div,\min,\max$, and every operation must be defined throughout $D$. The solutions are the fixed points of $F$. The class \emph{$\FIXP$} consists of the real-valued search problems that $\SL$-reduce to such problems, using the reductions defined next.
\end{definition}
The same class is obtained if the domain $D$ is restricted to a unit cube~\cite{EY10}.

\begin{definition}[Separable-linear reduction]\label{def:sl}
Let $\Pi$ and $\Gamma$ be search problems with real-valued solution sets. An \emph{$\SL$-reduction from $\Pi$ to $\Gamma$} consists of a polynomial-time map taking each instance $I$ of $\Pi$ to an instance $J=J(I)$ of $\Gamma$, together with indices $\rho(i)$ and rationals $a_i,b_i$ computable in polynomial time from $I$ such that every $\bm{y}\in\Sol_\Gamma(J)$ gives a vector $\bm{x}\in\Sol_\Pi(I)$ defined by
\[
 x_i=a_i y_{\rho(i)}+b_i.
\]
\end{definition}
This is the standard reduction used for $\FIXP$~\cite{EY10}.

A search problem $\Pi$ is $\FIXP$-complete if $\Pi\in\FIXP$ and every problem in $\FIXP$ admits an $\SL$-reduction to $\Pi$.

We use the following continuous-gate formulation of generalized circuits, based on~\cite[Definition~5.3]{FGHLP23}.
\begin{definition}[Generalized circuit]\label{def:generalized-circuit}
Let $\mathcal F$ be a family of continuous gate functions $f:[0,1]^k\to[0,1]$, each with a finite encoding. The nonnegative integer $k$ is the gate's \emph{arity}, meaning its number of inputs. For $N\geq1$, an \emph{$\mathcal F$-generalized circuit} is a list $\Gc=(g_1,\ldots,g_N)$ of gates
\[
 g_i=(f_i,j_{i,1},\ldots,j_{i,k_i}),
\]
where $f_i\in\mathcal F$ has arity $k_i$ and $j_{i,r}\in[N]$ for every input position $r\in[k_i]$. The instance specifies the encoding of each $f_i$ and its input indices. A gate of arity zero is a \emph{constant gate}. Input indices may repeat or equal $i$, and the directed graph with edges $j_{i,r}\to i$ may contain cycles.

An \emph{assignment} is a vector $\bm{x}\in[0,1]^N$. For $i\in[N]$, it \emph{satisfies} gate $g_i$ if $x_i=f_i(x_{j_{i,1}},\ldots,x_{j_{i,k_i}})$. A \emph{solution} of $\Gc$ is an assignment satisfying every gate; write $\Sol(\Gc)$ for the solution set. The associated search problem takes $\Gc$ as input and asks for a solution.
\end{definition}
For each gate $i\in[N]$, write $I_i=\{j_{i,h}:h\in[k_i]\}$ for its set of input indices; $I_i=\varnothing$ for a constant gate. A \emph{readout} of $\Gc$ is an ordered pair $(j,i)\in[N]^2$ with $j\in I_i$, identified with the directed edge $j\to i$: gate $i$ reads variable $x_j$. Repeated occurrences of $x_j$ in gate $i$ give the same readout $(j,i)$. In all auction constructions below, we first apply \cref{lem:bounded-copy} to make each gate's input indices pairwise distinct, so each readout then corresponds to exactly one input position.

Each variable $x_i$ is the output of exactly one gate, namely $g_i$, and may be used as an input by multiple gates.
Every generalized circuit has a solution by Brouwer's fixed-point theorem.

Circuits consisting only of constant gates are solved by assigning each variable the constant value of its gate. We therefore assume that input circuits contain at least one nonconstant gate.

Our convention allows \emph{self-inputs} (a gate reading its own output), as in~\cite{Rub18}, whereas~\cite[Definition~5.3]{FGHLP23} requires pairwise distinct input and output indices. When the complement gate $u\mapsto1-u$ is available, these conventions are $\SL$-equivalent: replace each gate input by a fresh copy obtained from two successive complement gates and recover the original variables by projection; the reverse reduction is the identity. The preprocessing used in the auction reductions is given in \cref{lem:bounded-copy}.

For $a,b,c\in\Q$ and rational $\tau>0$, define the clipped affine gate $A_{a,b,c}:[0,1]^2\to[0,1]$ and the reciprocal gate $R_{\tau}:[0,1]\to[0,1]$ by
\begin{equation}\label{eq:gate-functions}
 A_{a,b,c}(u,v):=\clip{au+bv+c}\quad\text{and}\quad
 R_{\tau}(u):=\frac1{1+\tau u}.
\end{equation}
Let $\A$ denote the family of all such clipped affine gates. Arguments with zero coefficients may be omitted, so constants and unary gates are included.

Our reductions use the following restricted gate set.
\begin{lemma}[Restricted Generalized Circuits; {\cite[Proposition~5.6]{FGHLP23}}]\label{thm:source}
Let $\mathcal F$ consist of the gate functions
\[
 u\mapsto1-u,\qquad (u,v)\mapsto\clip{u+v},\quad\text{and}\quad u\mapsto u^2.
\]
Finding a solution to an $\mathcal F$-generalized circuit is $\FIXP$-complete.
\end{lemma}

Restricting the fixed-point formulation to \emph{piecewise-affine circuits}, which use rational affine operations and $\min,\max$, gives the class $\LFIXP=\PPAD$~\cite{EY10}. Solutions may be required to be rational and written with binary numerators and denominators.

\section{Nonlinear Gates for Generalized Circuits}\label{sec:framework}

\subsection{Reciprocal Gates}\label{subsec:reciprocal-gates}
We prove \cref{thm:intro-framework} for every fixed rational $\tau>0$.
We reduce from the $\FIXP$-complete problem of \cref{thm:source}: we implement its three gate functions $u\mapsto1-u$, $(u,v)\mapsto\clip{u+v}$, and $u\mapsto u^2$ using the gates in $\A\cup\{R_{\tau}\}$.
The first two are already clipped affine gates, so the main construction implements squaring while keeping every intermediate value in $[0,1]$.

\begin{lemma}[Squaring with a reciprocal]\label{lem:square}
For every rational $\tau>0$, the squaring function $x\mapsto x^2$ on $[0,1]$ can be computed exactly by an acyclic circuit consisting of four clipped affine gates and three $R_{\tau}$ gates. The number of gates is independent of $\tau$, and the circuit can be constructed in time polynomial in the bit length of $\tau$.
\end{lemma}
\begin{proof}

For an input $x\in[0,1]$, define the intermediate values $r_+,r_-,t$ and output $s$ by the equations below. Each clipping operation and each occurrence of $R_{\tau}$ is one gate, so the displayed compositions use four clipped affine gates and three reciprocal gates.

\begin{align*}
 &r_+=R_{\tau}\left(\clip*{\frac{1+x}{2}}\right) &\text{and}\quad & r_-=R_{\tau}\left(\clip*{\frac{1-x}{2}}\right),\\
 &t=\clip*{\frac{2+\tau}{4\tau}(r_++r_-)-\frac1\tau} &\text{and}\quad &
 s=\clip*{\frac{(2+\tau)^2}{\tau^2}\bigl(1-R_{\tau}(t)\bigr)}.
\end{align*}
The arguments $(1\pm x)/2$ lie in $[0,1]$, so their clipping has no effect. Moreover,
\[
\begin{aligned}
 &r_++r_-=\frac{4(2+\tau)}{(2+\tau)^2-\tau^2x^2},\\
 &t=\clip*{\frac{2+\tau}{4\tau}(r_++r_-)-\frac1\tau} =\clip*{\frac{\tau x^2}{(2+\tau)^2-\tau^2x^2}}=\frac{\tau x^2}{(2+\tau)^2-\tau^2x^2}.
\end{aligned}
\]
The denominator satisfies $(2+\tau)^2-\tau^2x^2\geq4(1+\tau)>0$, so
$0\leq\frac{\tau x^2}{(2+\tau)^2-\tau^2x^2}\leq\frac{\tau}{4(1+\tau)}<\frac14$; hence the clipping in the equation for $t$ has no effect.
Therefore $R_{\tau}(t)=\frac{(2+\tau)^2-\tau^2x^2}{(2+\tau)^2}$.
Substitution into the equation for $s$ gives $s=\clip{x^2}=x^2\in[0,1]$. All coefficients in these seven gates have bit length polynomial in the bit length of $\tau$.
\end{proof}

\begin{theorem}[Completeness with reciprocal gates]\label{thm:universal}
For every fixed rational $\tau>0$, finding a solution to an $(\A\cup\{R_{\tau}\})$-generalized circuit (\cref{def:generalized-circuit}) is $\FIXP$-complete.
\end{theorem}
\begin{proof}
For membership, let the input gates be $g_i=(f_i,j_{i,1},\ldots,j_{i,k_i})$ with $f_i\in\A\cup\{R_{\tau}\}$ for every $i\in[N]$, and define $F_{\Gc}:[0,1]^N\to[0,1]^N$ by
\[
 \bigl(F_{\Gc}(\bm{x})\bigr)_i
 =f_i(x_{j_{i,1}},\ldots,x_{j_{i,k_i}}).
\]
Each $f_i$ is a clipped affine function or $R_{\tau}$, and every reciprocal denominator satisfies $1+\tau x_j\geq1$. Thus $F_{\Gc}$ is a continuous self-map represented by a polynomial-size algebraic circuit. By definition,
\[
 \bm{x}\in\Sol(\Gc)\quad\Longleftrightarrow\quad F_{\Gc}(\bm{x})=\bm{x},
\]
so $([0,1]^N,F_{\Gc})$ is a basic $\FIXP$ instance (\cref{def:fixp}) with the same solution set.

For hardness, let $\Gc$ be an $N$-gate generalized circuit from \cref{thm:source}. The gates $x_i=1-x_j$ and $x_i=\clip{x_j+x_k}$ are already clipped affine gates. Replace each gate $x_i=x_j^2$ by a fresh copy of the seven-gate circuit from \cref{lem:square}, retaining $x_i$ as its output and introducing six fresh variables for its other gate outputs. Write $\Gc'$ for the resulting circuit.

For every fixed assignment $\bm{x}\in[0,1]^N$ to the original variables, the six fresh intermediate values replacing a gate $x_i=x_j^2$ are uniquely determined by acyclic evaluation on input $x_j$. By \cref{lem:square}, the remaining equation, whose output is the retained variable $x_i$, holds exactly when $x_i=x_j^2$. Thus every solution of $\Gc'$ restricts to a solution of $\Gc$. The fresh intermediate variables for different replaced gates are disjoint, so this conclusion also holds for cycles and for $i=j$. The new circuit has at most $7N$ gates and can be constructed in polynomial time. Restriction to the original coordinates is an $\SL$-reduction.
\end{proof}

\begin{remark}[The reciprocal parameter]
There is no upper bound on $\tau$. Also, whenever $0<\eta\leq\tau$,
$R_{\tau}((\eta/\tau)x)=R_{\eta}(x)$. The only excluded value is zero: $R_0(x)=1$ is already a clipped affine gate, so this case remains within the piecewise-affine setting characterized by $\LFIXP=\PPAD$~\cite{EY10}.
\end{remark}

\subsection{Inverse Gates}\label{subsec:reciprocal-extraction}
The purpose of this subsection is to replace the reciprocal gate in \cref{thm:universal} by gate functions that arise in the ROI and GSP auction constructions of \cref{sec:roi,sec:gsp}. Their equilibrium conditions enforce equations $y=G(z)$, where $G:[0,1]\to[0,1]$ is a continuous strictly decreasing bijection.

We implement the inverse function $G^{-1}$ as a \emph{pseudogate} over $\A\cup\{G\}$, following the idea of Filos-Ratsikas, Hansen, H\o gh, and Hollender~\cite[Section~3.2]{FRHH23}: the required output is guaranteed whenever the auxiliary variables satisfy their fixed-point equations.

\begin{lemma}[Inverse Gate]\label{lem:inverse}
Let $G:[0,1]\to[0,1]$ be a continuous strictly decreasing bijection. For each fixed input $x\in[0,1]$, the equations in the unknowns $y,z\in[0,1]$
\[
 y=G(z)\quad\text{and}\quad z=\clip{z+y-x}
\]
have the unique solution $y=x$ and $z=G^{-1}(x)$.
\end{lemma}
\begin{proof}
We first show that every solution has $y=x$, including at the clipping endpoints. If $0<z<1$, clipping cannot change the value in the second equation. Thus $z=z+y-x$, which gives $y=x$.

At the endpoints, the second equation and the values $G(0)=1$, $G(1)=0$ give
\[
\begin{aligned}
 z=0&\ \Longrightarrow\ 1=y\leq x\leq1,\\
 z=1&\ \Longrightarrow\ 0\leq x\leq y=0.
\end{aligned}
\]
Thus $y=x$ also holds at both endpoints.

In every case $G(z)=x$, so bijectivity gives $z=G^{-1}(x)$ uniquely. Conversely, $y=x$ and $z=G^{-1}(x)$ satisfy $G(z)=y$ and $\clip{z+y-x}=\clip{z}=z$, proving existence.
\end{proof}
The three-input expression in \cref{lem:inverse} can be written as $w=(z+y)/2$ and $z=\clip{2w-x}$. Both equations use two-input clipped affine gates, since $(z+y)/2\in[0,1]$. We can now recover the reciprocal by an affine operation.

\begin{corollary}[Reciprocal Gate Replacement]\label{cor:extraction}
Fix rationals $a,b,c,\tau$ with $c\ne0$ and $\tau>0$, and a continuous strictly decreasing bijection $G:[0,1]\to[0,1]$ satisfying, for every $u\in[0,1]$,
\[
 G^{-1}(u)=a+bu+cR_{\tau}(u).
\]
Finding a solution to an $(\A\cup\{G\})$-generalized circuit (\cref{def:generalized-circuit}) is $\FIXP$-complete.
\end{corollary}
\begin{proof}

For membership, we construct an algebraic self-map with the same solutions as the given generalized circuit. By hypothesis, $G^{-1}(u)=a+bu+cR_{\tau}(u)$ has an algebraic-circuit representation. Given an $N$-gate $(\A\cup\{G\})$-generalized circuit $\Gc$, write each clipped affine gate as $x_i=A_{a_i,b_i,c_i}(x_{j_i},x_{k_i})$ and each $G$-gate as $x_i=G(x_{j_i})$. Define $F_{\Gc}:[0,1]^N\to[0,1]^N$ by

\[
 \bigl(F_{\Gc}(\bm{x})\bigr)_i=
 \begin{cases}
 A_{a_i,b_i,c_i}(x_{j_i},x_{k_i}),&\text{for an affine gate},\\
 \clip[\big]{x_i+G^{-1}(x_i)-x_{j_i}},&\text{for a }G\text{-gate}.
 \end{cases}
\]

The inverse $G^{-1}$ is a continuous strictly decreasing bijection of $[0,1]$. Apply \cref{lem:inverse} with gate function $G^{-1}$, fixed input $x_{j_i}$, and unknowns $y=G^{-1}(x_i)$ and $z=x_i$. This gives

\[
 x_i=\clip[\big]{x_i+G^{-1}(x_i)-x_{j_i}}
 \quad\Longleftrightarrow\quad G^{-1}(x_i)=x_{j_i}
 \quad\Longleftrightarrow\quad x_i=G(x_{j_i}).
\]

Thus $\Fix(F_{\Gc})=\Sol(\Gc)$. The map $F_{\Gc}$ is continuous and has a polynomial-size algebraic circuit; every denominator is of the form $1+\tau x_i\geq1$. The identity recovery map therefore proves membership in $\FIXP$.

For the reduction, we implement $R_{\tau}$ for each fixed input. Replace each gate $v=R_{\tau}(x)$ by the following simultaneous gate equations:
\[
\begin{aligned}
 &y=G(z)\quad\text{and}\quad w=A_{1/2,1/2,0}(z,y),\\
 &z=A_{2,-1,0}(w,x)\quad\text{and}\quad v=A_{1/c,-b/c,-a/c}(z,x),
\end{aligned}
\]
using fresh $y,z,w$ for each occurrence. Since $w=(z+y)/2$, the first three equations are equivalent to those in \cref{lem:inverse}, so their unique solution for each fixed $x$ is
\[
 y=x,\qquad z=G^{-1}(x),\quad\text{and}\quad w=\frac{G^{-1}(x)+x}{2},
\]
and the last equation gives $v=R_{\tau}(x)\in[0,1]$. Thus every new solution projects to an original solution, including when $v$ and $x$ are the same variable. The replacement adds three variables per reciprocal gate and is polynomial time. Projection is an $\SL$-reduction, so completeness follows from \cref{thm:universal}.
\end{proof}

\subsection{A Three-Gate Basis}\label{subsec:finite-affine-basis}
To implement clipped affine gates, the auction reductions in
\cref{sec:roi,sec:gsp,sec:throttling} use only the constant gate $1$
and the gate
\begin{equation}\label{eq:native-aff}
 H\colon (u,v) \mapsto \min\{1,2-u-v\}.
\end{equation}

\begin{lemma}\label{lem:auction-affine}
Every clipped affine gate $A_{a,b,c}$ can be simulated exactly by a generalized circuit over $\{1,H\}$, constructible in time polynomial in the bit length of $(a,b,c)$. Consequently, for every fixed rational $\tau>0$, finding a solution to a $\{1,H,R_\tau\}$-generalized circuit is $\FIXP$-complete.
\end{lemma}
\begin{proof}
We implement the complement and sum gates using $1$ and $H$ as follows:
\[
 C(u)=H(1,u)=1-u,\qquad
 S(u,v)=H(C(u),C(v))=\clip{u+v}.
\]
Truncated subtraction can be implemented by $C(S(C(u),v))=[u-v]_+$.

We adapt the constructions of integer multiples and rational constants in~\cite[Appendix~E.2]{FGHLP23}. For a positive integer $p$, write its binary expansion as $p=\sum_{i=0}^{k}b_i2^i$, where $b_i\in\{0,1\}$. On input $y\in[0,1]$, repeated doubling with $S$ computes $y,\clip{2y},\ldots,\clip{2^ky}$. Adding the terms with $b_i=1$, again using $S$, gives
\[
 \clip*{\sum_{i:b_i=1}\clip{2^iy}}=\clip{py}.
\]
For $\xi,\eta\geq0$, the identity $\clip{\clip{\xi}+\clip{\eta}}=\clip{\xi+\eta}$ shows that each intermediate clipping preserves the final clipped sum $\clip{py}$. The construction uses $O(\log(p+1))$ gates; for $p=0$, use $C(1)=0$.

For an integer $q\geq2$ and a fixed input $u\in[0,1]$, impose the equation $y=[u-\clip{(q-1)y}]_+$ on $y\in[0,1]$, using the integer-multiple and truncated-subtraction constructions above. If $y=0$, this equation forces $u=0$. If $y>0$, then $\clip{(q-1)y}=u-y<1$, so $(q-1)y=u-y$ and $y=u/q$. Conversely, $y=u/q$ satisfies the equation because $0\leq(q-1)u/q<1$. Thus the solution is unique for every $u$. For a rational $r=p/q\in[0,1]$ with integers $p\geq0$ and $q\geq1$, first construct $y=u/q$ and then output $\clip{py}=ru$. Here $q=1$ needs no division, and $py\leq1$, so the final clipping has no effect. The construction takes time polynomial in the bit lengths of $p$ and $q$.

To cope with intermediate clipping during the computation of $(u,v)\mapsto\clip{au+bv+c}$, we normalize by a common integer $K$ before summing. Set $K=\lceil\max\{1,|a|+|b|+|c|\}\rceil$. To handle negative coefficients, use $a=[a]_+-[-a]_+$, and similarly for $b,c$, to separate the positive contributions from the magnitudes of the negative contributions. Using these scalings, $1$, and $S$, compute the two nonnegative sums
\[
 s_+=\frac{[a]_+u+[b]_+v+[c]_+}{K},\qquad
 s_-=\frac{[-a]_+u+[-b]_+v+[-c]_+}{K}.
\]
Each summand is nonnegative and
$0\leq s_+,s_-\leq(|a|+|b|+|c|)/K\leq1$, so clipping does not change any partial sum defining $s_+$ or $s_-$. Zero terms use $C(1)=0$. The truncated-subtraction construction outputs $[s_+-s_-]_+$; multiplying this value by $K$ with the integer-multiple construction outputs
$\clip{K[s_+-s_-]_+}=\clip{au+bv+c}=A_{a,b,c}(u,v)$.

The construction is polynomial time and exact for every fixed input. Fresh auxiliary variables therefore permit replacement also in cyclic circuits, with projection giving an $\SL$-reduction.
Since $1,H\in\A$, \cref{thm:universal} and this reduction prove the claimed $\FIXP$-completeness.
\end{proof}

\begin{remark}[Two-gate basis]\label{rem:two-gate-basis}
  For some values of $\tau$, $\FIXP$-completeness holds even without the constant gate. For example, finding a solution to an $\{H,R_1\}$-generalized circuit is $\FIXP$-complete. Indeed, the equations
  \[
   a=H(a,a),\quad b=R_1(a),\quad c=H(b,b),\quad d=H(c,c),\quad z=H(d,d)
  \]
  force $(a,b,c,d,z)=(\frac{2}{3},\frac{3}{5},\frac{4}{5},\frac{2}{5},1)$. Replacing each constant gate by a fresh copy with its original output coordinate as $z$ gives an $\SL$-reduction from the $\{1,H,R_1\}$-circuit problem. Completeness follows from \cref{lem:auction-affine}.

  On the other hand, such a simulation of the constant gate is not possible for arbitrary $\tau$. At $\tau=3/4$, both gates fix $2/3$, so every $\{H,R_{3/4}\}$-generalized circuit has the solution $x_i=2/3$ for all $i$. In particular, these gates cannot force a variable to equal $1$.
\end{remark}

We also use the following preprocessing. The ``Moreover'' part in the following lemma is used only in \cref{sec:gsp}.

\begin{lemma}\label{lem:bounded-copy}
Let $\mathcal F$ contain $1,H$, and suppose every gate has at most two inputs. In polynomial time, an $N$-gate $\mathcal F$-generalized circuit $\Gc$ can be replaced by an $N'$-gate circuit $\Gc'$ over the same family with $N'= O(N)$ and satisfying the following properties:
\begin{itemize}
 \item For each $j\in[N']$, there are at most two readouts $(j,i)$ of $\Gc'$ if $x'_j$ is nonconstant, and at most one if $x'_j$ is constant.
 \item Each gate's input and output indices are pairwise distinct.
\end{itemize}
The variables $\{x'_1,\dots,x'_{N'}\}$ of $\Gc'$ contain the original variables $\{x_1,\dots,x_N\}$ of $\Gc$ as a subset, and a solution of $\Gc$ can be recovered from any solution of $\Gc'$ by projection.

Moreover, in time $\mathrm{poly}(N')$, we can find an assignment $\mathsf{A}\colon \{x'_1,\dots,x'_{N'}\}\to \{0,1,2\}$ of types to variables of $\Gc'$ with the following properties:
\begin{itemize}
  \item For every nonconstant gate of the form $x'_i=f(x'_j)$, it holds that $\mathsf{A}(x'_i)=0$ and $\mathsf{A}(x'_j)=1$.
  \item For every nonconstant gate of the form $x'_i=f(x'_j,x'_k)$ with $j\ne k$, the types $\mathsf{A}(x'_i),\mathsf{A}(x'_j),\mathsf{A}(x'_k)$ are distinct.
 \item If a variable $x'_i$ is in the original variables of $\Gc$, then $\mathsf{A}(x'_i)=0$.
 \end{itemize}
\end{lemma}
\begin{proof}
We construct $\Gc'$ and $\mathsf{A}$ together as follows.

Retain every original variable $x_i$ and set $\mathsf{A}(x_i)=0$ for $i\in[N]$.
For each nonconstant variable $x$ of $\Gc$, enumerate its $r$ input occurrences and set $x^{(0)}=x$. For $s=1,\ldots,r$ in order, let $h\in\{1,2\}$ be the position of the $s$th occurrence within its original gate. Introduce fresh variables $y^{(s)},x^{(s)}$ with gates
\begin{equation}\label{eq:copy-gadget}
 y^{(s)}=H(1,x^{(s-1)}),\qquad x^{(s)}=H(1,y^{(s)}),
\end{equation}
using a fresh constant gate for each displayed $1$. Set $\mathsf{A}(x^{(s)})=h$ and choose
\[
 \mathsf{A}(y^{(s)})\in\{0,1,2\}\setminus\{\mathsf{A}(x^{(s-1)}),\mathsf{A}(x^{(s)})\}.
\]
In each of the two $H$-gates in \eqref{eq:copy-gadget}, the other input and the output now have distinct types, so assign its fresh constant variable the unique remaining type in $\{0,1,2\}$. Replace the $s$th input occurrence of $x$ by $x^{(s)}$.

For each occurrence of an original constant variable at input position $h\in\{1,2\}$, use a fresh copy of its defining constant gate and assign its output type $h$. All introduced variables are distinct from one another and from the original variables, so no type is assigned twice. This defines $\Gc'$ and $\mathsf{A}$ on all its variables. The output and input types at the modified original gates are $(0,1)$ for unary gates and $(0,1,2)$ for binary gates. Every added nonconstant gate is one of the $H$-gates in \eqref{eq:copy-gadget}, whose types are pairwise distinct by construction. Together with $\mathsf{A}(x_i)=0$ for $i\in[N]$, this proves the moreover part.

Since $H(1,u)=1-u$ for $u\in[0,1]$, \eqref{eq:copy-gadget} gives $x^{(s)}=x^{(s-1)}=x$ for all $s\in[r]$. Thus every solution of $\Gc'$ restricts to a solution of $\Gc$, including when $\Gc$ has cycles or self-inputs. Each original nonconstant variable is used only in the gate defining $y^{(1)}$, if it has any input occurrences. Each $x^{(s)}$ is used at its assigned original input occurrence and, for $s<r$, in the gate defining $y^{(s+1)}$. Each $y^{(s)}$ and each new constant variable is used once, while original constant variables are unused. All gate inputs and outputs have distinct indices by freshness. Each original input occurrence introduces at most four gates, and there are at most $2N$ such occurrences, so $N'=O(N)$. The construction and type assignment take $O(N')$ time.
\end{proof}

\section{Pacing Equilibria in ROI-constrained Auctions}\label{sec:roi}
We prove \cref{thm:intro-roi} for the second-price ROI pacing model of Li and Tang~\cite{LT24}, using the formulation of Filos-Ratsikas, Hansen, H\o gh, and Hollender~\cite[Section~7.2]{FRHH24}.

\subsection{Model}
Fix a rational $\delta>0$. An instance specifies $n\geq2$ bidders, $m\geq1$ items, and nonnegative rational values $\bm{v}=(v_{\ell,s})_{\ell\in[n],s\in[m]}$. Each item $s\in[m]$ has a positive value to at least one bidder $\ell\in[n]$. Bidder $\ell$ chooses a multiplier $\alpha_\ell\in[1,1+\delta]$ and bids $\alpha_\ell v_{\ell,s}$ on item $s$. For $s\in[m]$ and a multiplier vector $\bm{\alpha}=(\alpha_\ell)_{\ell\in[n]}\in[1,1+\delta]^n$, let $h_s(\bm{\alpha})$ be the highest bid and let the price $p_s(\bm{\alpha})$ be the second-highest bid, counting tied bids separately. A fractional allocation is a matrix $\bm{z}\in[0,1]^{n\times m}$, where $z_{\ell,s}$ is the share of item $s$ allocated to bidder $\ell$.
\begin{definition}[Second-price ROI pacing equilibrium; {\cite[Definition~7.4]{FRHH24}}]\label{def:roi-pacing}
A pair $(\bm{\alpha},\bm{z})\in[1,1+\delta]^n\times[0,1]^{n\times m}$ is a \emph{second-price ROI pacing equilibrium} if the following conditions hold for all bidders $\ell\in[n]$ and items $s\in[m]$:
\begin{align}
 &\sum_{\ell=1}^n z_{\ell,s}=1,
  \label{eq:roi-full-allocation}\\
 &z_{\ell,s}>0\ \Longrightarrow\ \alpha_\ell v_{\ell,s}=h_s(\bm{\alpha}),
  \label{eq:roi-highest-bid-allocation}\\
 &\sum_{s=1}^m z_{\ell,s}p_s(\bm{\alpha})\leq\sum_{s=1}^m z_{\ell,s}v_{\ell,s},
  \label{eq:roi-feasibility}\\
 &\sum_{s=1}^m z_{\ell,s}p_s(\bm{\alpha})<\sum_{s=1}^m z_{\ell,s}v_{\ell,s}
  \ \Longrightarrow\ \alpha_\ell=1+\delta.
  \label{eq:roi-cap}
\end{align}
\end{definition}
The first two conditions allocate each item in full among its highest bidders; a tied bidder need not receive a positive share. The last two conditions require each bidder's total payment to be at most the total value of its allocation, with equality whenever its multiplier is below $1+\delta$.
\begin{theorem}[ROI-constrained pacing]\label{thm:roi-complete}
For every fixed rational $\delta>0$, the following search problem is $\FIXP$-complete. Given integers $n\geq2$, $m\geq1$, and binary-encoded nonnegative rational values $\bm{v}=(v_{\ell,s})_{\ell\in[n],s\in[m]}$ with $\max_{\ell\in[n]}v_{\ell,s}>0$ for every $s\in[m]$, find an exact second-price ROI pacing equilibrium $(\bm{\alpha},\bm{z})\in[1,1+\delta]^n\times[0,1]^{n\times m}$.
\end{theorem}
Membership in $\FIXP$ holds for every fixed rational $\delta>0$ by \cite[Theorem~7.3]{FRHH24}. We prove $\FIXP$-hardness in the remainder of this section.

\subsection{Gate Construction}\label{subsec:roi-construction}
Define $G_{\mathrm{ROI}}:[0,1]\to[0,1]$ by its inverse
\begin{equation}\label{eq:roi-inverse}
 G_{\mathrm{ROI}}^{-1}(x)=2-\frac43x-R_{1/2}(x),
\end{equation}
where $R_{1/2}(x)=1/(1+x/2)$. By \cref{lem:roi-gate}, $G_{\mathrm{ROI}}$ is a continuous strictly decreasing bijection, and the $\{1,H,G_{\mathrm{ROI}}\}$-generalized circuit problem is $\FIXP$-complete.

Let $\Gc$ be such a circuit. Apply \cref{lem:bounded-copy} to make each gate's input and output indices pairwise distinct. Write $N$ for the resulting number of gates.

Set $L=1+2\delta/3$. Each nonconstant gate $i$ has two bidders $A_i,B_i$ with the intended encoding
\begin{equation}\label{eq:roi-encoding}
 \alpha_{A_i}=L+\frac{\delta}{3}x_i.
\end{equation}
Introduce one shared bidder $\mathsf{One}$ and one private item $q_{\mathsf{One}}$ valued at $1$ only by this bidder. For every constant gate $x_i=1$, set $A_i=\mathsf{One}$; no bidder $B_i$ is needed.

For every nonconstant gate $i$, introduce one controller item $s_i$ and one private item $q_{A_i}$. For each readout $(j,i)$, introduce one item $r_{ji}$. For every $H$-gate, introduce an additional private item $q_{B_i}$. All these items are fresh, including the two readout items when both inputs are constants.

Let $\mathcal J$ be the set of these items. Together with the bidders above, they form the auction instance $\mathcal I_{\Gc}$. The nonzero values of the gate items are specified in \cref{tab:roi-valuations}; all unlisted values are zero.
\begin{table}[H]
\centering
\renewcommand{\arraystretch}{1.25}
\begin{tabular}{llccc}
\toprule
Gate & Item & $A_i$ & $B_i$ & $A_j$ \\
\midrule
$H$-gate $i$ & Controller $s_i$ & $1$ & $1$ & $0$ \\
& Private $q_{A_i}$ & $\delta/3$ & $0$ & $0$ \\
& Private $q_{B_i}$ & $0$ & $\delta/3$ & $0$ \\
& Readout item $r_{ji}$ & $1+\delta$ & $0$ & $1$ \\
\midrule
$G_{\mathrm{ROI}}$-gate $i$ & Controller $s_i$ & $1$ & $1+\delta/3$ & $0$ \\
& Private $q_{A_i}$ & $\delta/2$ & $0$ & $0$ \\
& Readout item $r_{ji}$ & $0$ & $3/4+5\delta/6$ & $3/4$ \\
\bottomrule
\end{tabular}
\caption{Values of the items associated with nonconstant gate $i$. Each entry is the column bidder's value for the row item. There is one item $r_{ji}$ for each readout $(j,i)$; $A_j=\mathsf{One}$ for a constant input.}
\label{tab:roi-valuations}
\end{table}

\subsection{Correctness}
The following three lemmas concern the full auction $\mathcal I_{\Gc}$ constructed in \cref{subsec:roi-construction}. For an equilibrium $(\bm{\alpha},\bm{z})$ and each bidder $\ell$, define its surplus (total value minus total payment) by
\[
 \Delta_\ell=\sum_{s\in\mathcal J}z_{\ell,s}\bigl(v_{\ell,s}-p_s(\bm{\alpha})\bigr).
\]
By \eqref{eq:roi-feasibility} and \eqref{eq:roi-cap}, $\Delta_\ell\geq0$ and $\alpha_\ell<1+\delta\Longrightarrow\Delta_\ell=0$.

\begin{lemma}[Readout allocation]\label{lem:roi-readout}
In every equilibrium $(\bm{\alpha},\bm{z})$ of $\mathcal I_{\Gc}$, we have $\alpha_{\mathsf{One}}=1+\delta$ and $\alpha_{A_i}>1$ for all $i\in[N]$. Each readout item $r_{ji}$ is allocated entirely to $A_i$ if gate $i$ is an $H$-gate and to $B_i$ otherwise. Its price is
\[
 p_{r_{ji}}(\bm{\alpha})=
 \begin{cases}
 \alpha_{A_j},&\text{for an }H\text{-gate},\\
 \frac34\alpha_{A_j},&\text{for a }G_{\mathrm{ROI}}\text{-gate}.
 \end{cases}
\]
The input bidder $A_j$ receives no allocation and pays zero on $r_{ji}$.
\end{lemma}
\begin{proof}
Fix an equilibrium $(\bm{\alpha},\bm{z})$. For every nonconstant gate $i$, we have $\alpha_{A_i}>1$. Indeed, if $\alpha_{A_i}=1$, then for every $s\in\mathcal J$ with $z_{A_i,s}>0$, \eqref{eq:roi-highest-bid-allocation} gives $p_s(\bm{\alpha})\leq h_s(\bm{\alpha})=v_{A_i,s}$. Moreover, $z_{A_i,q_{A_i}}=1$ and $p_{q_{A_i}}(\bm{\alpha})=0$, so $\Delta_{A_i}\geq v_{A_i,q_{A_i}}\geq\frac{\delta}{3}>0$.
Equation~\eqref{eq:roi-cap} would force $\alpha_{A_i}=1+\delta>1$, a contradiction.

Using $\alpha_{B_i}\geq1$ and $\alpha_{A_j}\leq1+\delta$, the bids on $r_{ji}$ satisfy, for an $H$-gate,
\[
 (1+\delta)\alpha_{A_i}>1+\delta\ge\alpha_{A_j},
\]
and, for a $G_{\mathrm{ROI}}$-gate,
\[
 \left(\frac34+\frac{5\delta}{6}\right)\alpha_{B_i}>\frac34 + \frac{3\delta}{4}\ge\frac34\alpha_{A_j}.
\]
Thus $z_{A_i,r_{ji}}=1$ and $p_{r_{ji}}(\bm{\alpha})=\alpha_{A_j}$ for an $H$-gate, whereas $z_{B_i,r_{ji}}=1$ and $p_{r_{ji}}(\bm{\alpha})=\tfrac34\alpha_{A_j}$ for a $G_{\mathrm{ROI}}$-gate. In both cases $z_{A_j,r_{ji}}=0$. In particular, $\mathsf{One}$ loses every readout item on which it bids. Due to the private item $q_{\mathsf{One}}$, we have $\Delta_{\mathsf{One}}=1>0$, so \eqref{eq:roi-cap} forces $\alpha_{\mathsf{One}}=1+\delta$.
\end{proof}

\begin{lemma}[$H$-gate equations]\label{lem:roi-affine}
In every equilibrium $(\bm{\alpha},\bm{z})$ of $\mathcal I_{\Gc}$, each $H$-gate $i$ satisfies
\[
 \alpha_{A_i}=\alpha_{B_i}
 =\min\left\{1+\delta,L+\sum_{j\in I_i}(1+\delta-\alpha_{A_j})\right\}\in[L,1+\delta].
\]
\end{lemma}
\begin{proof}
Fix an equilibrium $(\bm{\alpha},\bm{z})$ and an $H$-gate $i$. Only $A_i$ and $B_i$ have positive values for $s_i$, so $z_{A_i,s_i}+z_{B_i,s_i}=1$. The private items $q_{A_i}$ and $q_{B_i}$ contribute $\delta/3$ to $\Delta_{A_i}$ and $\Delta_{B_i}$, respectively. By \cref{lem:roi-readout}, the items $r_{ji}$ with $j\in I_i$ contribute $\sum_{j\in I_i}(1+\delta-\alpha_{A_j})\geq0$ to $\Delta_{A_i}$ and zero to $\Delta_{B_i}$. For every item $r$ on which $A_i$ is an input bidder, \cref{lem:roi-readout} gives $z_{A_i,r}=0$, so these items contribute zero to $\Delta_{A_i}$. If $\alpha_{A_i}\ne\alpha_{B_i}$, the bidder $\ell\in\{A_i,B_i\}$ with the lower bid on $s_i$ would receive none of it and have $\Delta_\ell\geq\delta/3>0$. Equation~\eqref{eq:roi-cap} would force $\alpha_\ell=1+\delta$, contradicting its lower bid. Thus $\alpha_{A_i}=\alpha_{B_i}$ and $p_{s_i}(\bm{\alpha})=\alpha_{A_i}$.

Using $z_{A_i,s_i}+z_{B_i,s_i}=1$, $z_{A_i,q_{A_i}}=z_{B_i,q_{B_i}}=z_{A_i,r_{ji}}=1$ for $j\in I_i$, and $L=1+2\delta/3$, we obtain
\[
\begin{aligned}
 \Delta_{A_i}+\Delta_{B_i}
 &=\underbrace{(z_{A_i,s_i}+z_{B_i,s_i})\bigl(1-p_{s_i}(\bm{\alpha})\bigr)}_{s_i}
 +\underbrace{\frac{\delta}{3}\bigl(z_{A_i,q_{A_i}}+z_{B_i,q_{B_i}}\bigr)}_{q_{A_i},q_{B_i}}+\underbrace{\sum_{j\in I_i}z_{A_i,r_{ji}}\bigl(1+\delta-p_{r_{ji}}(\bm{\alpha})\bigr)}_{r_{ji}\ (j\in I_i)}\\
 &=1-\alpha_{A_i}+\frac{2\delta}{3}+\sum_{j\in I_i}(1+\delta-\alpha_{A_j})\\
 &=L+\sum_{j\in I_i}(1+\delta-\alpha_{A_j})-\alpha_{A_i} \ge 0.
\end{aligned}
\]
If $\alpha_{A_i}<1+\delta$, then $\Delta_{A_i}=\Delta_{B_i}=0$ and the identity above gives
$\alpha_{A_i}=L+\sum_{j\in I_i}(1+\delta-\alpha_{A_j})$.
If $\alpha_{A_i}=1+\delta$, the nonnegativity of $\Delta_{A_i}+\Delta_{B_i}$ instead gives
$1+\delta\leq L+\sum_{j\in I_i}(1+\delta-\alpha_{A_j})$.
These two cases prove the minimum formula. Since $\alpha_{A_j}\leq1+\delta$ for every $j\in I_i$, both arguments of the minimum are at least $L$.
\end{proof}

Define $\Psi_\delta\colon[L,1+\delta]\to[\delta/12,\delta/3]$ by
\begin{equation}\label{eq:roi-psi}
 \Psi_\delta(t)=\left(t-1-\frac{\delta}{3}\right)\left(1-\frac{\delta}{2(t-1)}\right).
\end{equation}

\begin{lemma}[$G_{\mathrm{ROI}}$-gate equations]\label{lem:roi-nonlinear}
In every equilibrium $(\bm{\alpha},\bm{z})$ of $\mathcal I_{\Gc}$, each $G_{\mathrm{ROI}}$-gate $x_i=G_{\mathrm{ROI}}(x_j)$ satisfies $\alpha_{A_i}\in[L,1+\delta]$ and
\[
 \Psi_\delta(\alpha_{A_i})=\frac{\delta}{3}-\frac34(\alpha_{A_j}-L).
\]
\end{lemma}
\begin{proof}
Fix an equilibrium $(\bm{\alpha},\bm{z})$ and consider a $G_{\mathrm{ROI}}$-gate $x_i=G_{\mathrm{ROI}}(x_j)$.

By \cref{lem:roi-readout}, among items other than $s_i$, only the private item $q_{A_i}$ contributes to $\Delta_{A_i}$, and only the readout item $r_{ji}$ contributes to $\Delta_{B_i}$. Using the values in \cref{tab:roi-valuations}, we obtain
\begin{equation}\label{eq:roi-nonlinear-available}
\begin{aligned}
 \sum_{s\in\mathcal J\setminus\{s_i\}}z_{A_i,s}(v_{A_i,s}-p_s(\bm{\alpha}))&=\frac{\delta}{2},\\
 \sum_{s\in\mathcal J\setminus\{s_i\}}z_{B_i,s}(v_{B_i,s}-p_s(\bm{\alpha}))
 &=\frac{\delta}{3}-\frac34(\alpha_{A_j}-L)\geq\frac{\delta}{12}.
\end{aligned}
\end{equation}
Now we have $z_{B_i,s_i}>0$: otherwise $\Delta_{B_i}\geq\delta/12>0$, so \eqref{eq:roi-cap} would force $\alpha_{B_i}=1+\delta$, giving the bid $(1+\delta/3)(1+\delta)>1+\delta\geq\alpha_{A_i}$ on $s_i$. Thus $z_{B_i,s_i}=1$ by \eqref{eq:roi-full-allocation}--\eqref{eq:roi-highest-bid-allocation}, contradicting $z_{B_i,s_i}=0$.

First suppose $\alpha_{A_i}<1+\delta$. Then \eqref{eq:roi-cap} implies $\Delta_{A_i}=0$, so \eqref{eq:roi-nonlinear-available} forces $z_{A_i,s_i}>0$. The bids for $s_i$ then tie, so $\alpha_{A_i}=(1+\delta/3)\alpha_{B_i}$ and $\alpha_{A_i},\alpha_{B_i}<1+\delta$. The second-price rule therefore gives $p_{s_i}(\bm{\alpha})=(1+\delta/3)\alpha_{B_i}=\alpha_{A_i}$. Using $z_{B_i,s_i}=1-z_{A_i,s_i}$, we obtain
\begin{equation}\label{eq:roi-controller}
\begin{aligned}
 \Delta_{A_i}&=\frac{\delta}{2}+z_{A_i,s_i}(1-\alpha_{A_i})=0,\\
 \Delta_{B_i}&=\frac{\delta}{3}-\frac34(\alpha_{A_j}-L)
 +(1-z_{A_i,s_i})\left(1+\frac{\delta}{3}-\alpha_{A_i}\right)=0.
\end{aligned}
\end{equation}
The equation for $\Delta_{A_i}$ in \eqref{eq:roi-controller} and $z_{B_i,s_i}>0$ give
$z_{A_i,s_i}=\delta/(2(\alpha_{A_i}-1))<1$, so $\alpha_{A_i}>1+\delta/2$. Substituting this value of $z_{A_i,s_i}$ into the equation for $\Delta_{B_i}$ and subtracting $\delta/12$ gives
\[
 \frac{(\alpha_{A_i}-L)(\alpha_{A_i}-1-\delta/4)}{\alpha_{A_i}-1}
 =\frac34(1+\delta-\alpha_{A_j})\geq0.
\]
Since $\alpha_{A_i}-1-\delta/4>0$ and $\alpha_{A_i}-1>0$, this implies $\alpha_{A_i}\geq L$. By \eqref{eq:roi-psi}, the left-hand side is now $\Psi_\delta(\alpha_{A_i})-\delta/12$, so the same equality gives $\Psi_\delta(\alpha_{A_i})=\delta/3-\frac34(\alpha_{A_j}-L)$.

For $\alpha_{A_i}=1+\delta$, the lower bound $\alpha_{A_i}\geq L$ is immediate. Together with \cref{lem:roi-readout,lem:roi-affine}, this proves $\alpha_{A_k}\in[L,1+\delta]$ for all $k\in[N]$. It remains to prove the identity for a capped output.

Fix a $G_{\mathrm{ROI}}$-gate $x_i=G_{\mathrm{ROI}}(x_j)$ with $\alpha_{A_i}=1+\delta$. The bound $\alpha_{A_j}\geq L$ gives $\delta/3-\frac34(\alpha_{A_j}-L)\leq\delta/3$. Suppose $z_{A_i,s_i}=0$. Then $z_{B_i,s_i}=1$ and $p_{s_i}(\bm{\alpha})=1+\delta$, giving
\[
 \Delta_{B_i}
 =-\frac{\delta}{3}-\frac34(\alpha_{A_j}-L)\leq-\frac{\delta}{3}<0,
\]
a contradiction. Thus $z_{A_i,s_i},z_{B_i,s_i}>0$, so \eqref{eq:roi-highest-bid-allocation} gives $1+\delta=\alpha_{A_i}=(1+\delta/3)\alpha_{B_i}$. Hence $\alpha_{B_i}=(1+\delta)/(1+\delta/3)<1+\delta$ and thus $\Delta_{B_i}=0$ by \eqref{eq:roi-feasibility} and \eqref{eq:roi-cap}.
Substituting $\alpha_{A_i}=1+\delta$ into the surplus expressions in \eqref{eq:roi-controller}, and using $\Delta_{A_i}\geq0$, $\Delta_{B_i}=0$, and $\alpha_{A_j}\geq L$, gives
\[
 \delta z_{A_i,s_i}\leq\frac{\delta}{2}\quad\text{and}\quad
 \frac{2\delta}{3}(1-z_{A_i,s_i})=\frac{\delta}{3}-\frac34(\alpha_{A_j}-L)\leq\frac{\delta}{3}.
\]
Thus $z_{A_i,s_i}=1/2$ and $\alpha_{A_j}=L$. Since $\Psi_\delta(1+\delta)=\delta/3$, the asserted identity follows also in this case.
\end{proof}

\begin{proof}[Proof of \cref{thm:roi-complete}]
Given an $N_0$-gate $\{1,H,G_{\mathrm{ROI}}\}$-generalized circuit $\Gc_0$, let $\Gc$ be its preprocessed circuit, retaining the original variables at indices $[N_0]$. Let $\mathcal I_{\Gc}$ be the second-price ROI pacing instance constructed in \cref{subsec:roi-construction}.

Let $(\bm{\alpha},\bm{z})$ be any equilibrium of $\mathcal I_{\Gc}$, whose existence follows from \cite[Theorem~7.2]{FRHH24}. By \cref{lem:roi-readout,lem:roi-affine,lem:roi-nonlinear}, every $\alpha_{A_i}$ lies in $[L,1+\delta]$. For each $i\in[N]$, define
\begin{equation}\label{eq:roi-recovery}
 x_i=\frac{3(\alpha_{A_i}-L)}{\delta}\in[0,1].
\end{equation}
For every constant gate, $A_i=\mathsf{One}$ and $\alpha_{A_i}=1+\delta$ by \cref{lem:roi-readout}, so \eqref{eq:roi-recovery} gives $x_i=1$. For each $H$-gate $x_i=H(x_j,x_k)$, substitute $I_i=\{j,k\}$, $\alpha_{A_\ell}=L+\frac{\delta}{3}x_\ell$ for $\ell\in\{i,j,k\}$, and $1+\delta=L+\delta/3$ into \cref{lem:roi-affine}. Dividing by $\delta/3$ after subtracting $L$ gives
\[
 x_i=\min\{1,2-x_j-x_k\}=H(x_j,x_k).
\]
For each gate $x_i=G_{\mathrm{ROI}}(x_j)$, substituting $\alpha_{A_\ell}=L+\frac{\delta}{3}x_\ell$ for $\ell\in\{i,j\}$ into \cref{lem:roi-nonlinear} gives
\[
 \Psi_\delta\left(L+\frac{\delta}{3}x_i\right)=\frac{\delta}{3}\left(1-\frac34x_j\right).
\]
For every $x\in[0,1]$, direct substitution in \eqref{eq:roi-psi} and \eqref{eq:roi-inverse} gives
\[
 \frac{3}{\delta}\Psi_\delta\left(L+\frac{\delta}{3}x\right)
 =x-\frac12+\frac{3}{2(2+x)}
 =1-\frac34G_{\mathrm{ROI}}^{-1}(x).
\]
Applying this identity with $x=x_i$ gives $G_{\mathrm{ROI}}^{-1}(x_i)=x_j$, hence $x_i=G_{\mathrm{ROI}}(x_j)$, including at the endpoints. Thus $\bm{x}\in\Sol(\Gc)$. The lemmas apply to the full auction, including when the circuit contains directed cycles. By \cref{lem:bounded-copy}, restriction to indices $[N_0]$ gives $(x_i)_{i\in[N_0]}\in\Sol(\Gc_0)$.

For every fixed rational $\delta>0$, the construction has linearly many items and bidders and is computable in polynomial time. All specified positive values are positive rationals of constant bit length. The recovery in \eqref{eq:roi-recovery} gives an $\SL$-reduction. Together with membership in $\FIXP$ from~\cite[Theorem~7.3]{FRHH24}, it proves the theorem.
\end{proof}

\section{Autobidding in Sponsored Shopping}\label{sec:gsp}
We prove \cref{thm:intro-gsp} in the sponsored-shopping model of~\cite{DLPST26}, using the budget equilibrium notion of~\cite{HLN26}. See~\cite{EOS07} for classical GSP.

\subsection{Model}\label{subsec:gsp-model}
An instance specifies a bidder set $\mathcal B$, an item set $\mathcal J$, and ownership sets $(\mathcal J_{\ell})_{\ell\in\mathcal B}$ that partition $\mathcal J$. Each item is a product whose owner bids to advertise it. Item $s$ has a positive rational value $v_s$ to its owner, and bidder $\ell$ has a positive rational budget $M_{\ell}$. We consider one advertising slot per item and click-through rate one at every slot. Thus all items are displayed, and each item's payment equals its price.

Each bidder $\ell$ uses one multiplier $\alpha_{\ell}\in[0,1]$ for all its items, giving the bids $b_s=\alpha_{\ell} v_s$ for $s\in\mathcal J_{\ell}$.

Write $m=|\mathcal J|$. A ranking $\sigma=(s_1,\ldots,s_m)$ is a permutation of $\mathcal J$, with item $s_k$ displayed in slot $k$. For $s_k\in\mathcal J_{\ell}$, define its price by
\begin{equation}\label{eq:gsp-item-price}
 c_k(\bm{\alpha},\sigma)
 =\max\!\left(\{b_{s_{k'}}:k<k'\leq m,\ s_{k'}\notin\mathcal J_{\ell}\}\cup\{0\}\right).
\end{equation}
For a ranking $\sigma$ in nonincreasing order of bids, $c_k(\bm{\alpha},\sigma)$ is the first lower-ranked bid belonging to a different bidder, as required by the \emph{generalized second-price} (GSP) rule; items of the same owner are skipped. The price is zero if no such item exists. \Cref{tab:gsp-pricing-example} illustrates the rule: $s_2$ is skipped when pricing $s_1$ because both items belong to bidder $A$.

\begin{table}[htbp]
\centering
\small
\begin{tabular}{cccccc}
\toprule
Slot & Item & Bidder & Bid $b_s$ & Item setting the price & Price \\
\midrule
$1$ & $s_1$ & $A$ & $10$ & $s_3$ & $6$ \\
$2$ & $s_2$ & $A$ & $8$  & $s_3$ & $6$ \\
$3$ & $s_3$ & $B$ & $6$  & $s_4$ & $4$ \\
$4$ & $s_4$ & $A$ & $4$  & None  & $0$ \\
\bottomrule
\end{tabular}
\caption{An example of GSP pricing with bids.}
\label{tab:gsp-pricing-example}
\end{table}

To specify how ties are resolved, let $\pi$ be a probability distribution over rankings. Bidder $\ell$'s expected total payment, averaging over $\pi$, is
\begin{equation}\label{eq:gsp-total-payment}
 P_{\ell}(\bm{\alpha},\pi)
 =\mathbb E_{\sigma\sim\pi}\!\left[
   \sum_{\substack{k\in[m]\\s_k\in\mathcal J_{\ell}}}
   c_k(\bm{\alpha},\sigma)
 \right].
\end{equation}

\begin{definition}[Budget-constrained autobidding equilibrium; {\cite[Definition~2.2]{HLN26}}]\label{def:abe}
A pair $(\bm{\alpha},\pi)$, where $\bm{\alpha}\in[0,1]^{\mathcal B}$ and $\pi$ is a probability distribution over rankings of $\mathcal J$, is a \emph{budget-constrained autobidding equilibrium} if the following conditions hold for every ranking $\sigma=(s_1,\ldots,s_m)$ of $\mathcal J$ and every bidder $\ell\in\mathcal B$:
\begin{align}
 &\pi(\sigma)>0\ \Longrightarrow\ b_{s_1}\geq\cdots\geq b_{s_m},
  \label{eq:gsp-ranking}\\
 &P_{\ell}(\bm{\alpha},\pi)\leq M_{\ell},
  \label{eq:gsp-budget}\\
 &P_{\ell}(\bm{\alpha},\pi)<M_{\ell}\ \Longrightarrow\ \alpha_{\ell}=1.
  \label{eq:gsp-unused-budget}
\end{align}
\end{definition}
In this section, ``equilibrium'' means a budget-constrained autobidding equilibrium.

By \eqref{eq:gsp-ranking}, only the relative order of equal bids can vary under $\pi$. Condition \eqref{eq:gsp-budget} bounds each bidder's expected total payment by its budget. Condition \eqref{eq:gsp-unused-budget} is the \emph{unused-budget condition}: a bidder with unused budget must use multiplier one. Together, \eqref{eq:gsp-budget} and \eqref{eq:gsp-unused-budget} imply
\begin{equation}\label{eq:gsp-tight-budget}
 \alpha_{\ell}<1\ \Longrightarrow\ P_{\ell}(\bm{\alpha},\pi)=M_{\ell}.
\end{equation}

Hansen, Li, and Nie~\cite{HLN26} prove that every instance has an equilibrium in which only polynomially many rankings have positive probability. Such a distribution has \emph{polynomial-size support}: it is specified by a list of these rankings and their probabilities.

\begin{theorem}[Autobidding]\label{thm:gsp-complete}
Given a GSP sponsored-shopping instance with one slot per item and all click-through rates equal to one, computing a multiplier vector $\bm{\alpha}\in[0,1]^{\mathcal B}$ of a budget-constrained autobidding equilibrium (\cref{def:abe}) is $\FIXP$-complete.% Computing a full equilibrium in the representation of~\cite{HLN26} is also $\FIXP$-complete.
\end{theorem}
Membership follows from~\cite[Theorem~1.1]{HLN26}: the positive rational values and budgets, multiplier interval $[0,1]$, and pricing and equilibrium conditions above are a special case of their model, with the number of slots equal to the number of items and every click-through rate equal to one. Projecting a full equilibrium onto its multiplier coordinates also gives membership for the multiplier-only problem. Hardness for this special case therefore establishes completeness for general slot counts and click-through rates as well.

\subsection{Gate Construction}\label{subsec:gsp-construction}
Define $G_{\mathrm{GSP}}:[0,1]\to[0,1]$ by its inverse
\begin{equation}\label{eq:gsp-ginv}
 G_{\mathrm{GSP}}^{-1}(x)=2-\frac54x-R_{1/3}(x),
\end{equation}
where $R_{1/3}(x)=1/(1+x/3)$. By \cref{lem:gsp-gate}, $G_{\mathrm{GSP}}$ is a continuous strictly decreasing bijection, and the $\{1,H,G_{\mathrm{GSP}}\}$-generalized circuit problem is $\FIXP$-complete.

Let $\Gc_0$ be such a circuit with $N_0$ gates. Apply \cref{lem:bounded-copy} to obtain a circuit $\Gc$ with $N=O(N_0)$ gates and types $t_i=\mathsf{A}(x_i)\in\{0,1,2\}$ for $i\in[N]$. The types satisfy $t_i=0$ for $i\in[N_0]$, $\{t_i,t_j,t_k\}=\{0,1,2\}$ for each gate $x_i=H(x_j,x_k)$, and $(t_i,t_j)=(0,1)$ for each gate $x_i=G_{\mathrm{GSP}}(x_j)$.

\paragraph{Bidders and Items.}
The bidder set $\mathcal B$ consists of two bidders $A_j^{(1)},A_j^{(2)}$ for each nonconstant gate, one bidder $Q_j$ for each constant gate $j$ with at least one readout $(j,i)$, and one shared bidder $\mathsf{Guard}$.

The set of all items is $\mathcal J = \bigsqcup_{i\text{ nonconstant}}\mathcal J_i^{\mathsf{ctl}}\sqcup\bigsqcup_{\substack{i\text{ nonconstant}\\j\in I_i}}\mathcal J_{ji}^{\mathsf{read}}$, where $\mathcal{J}_i^{\mathrm{ctl}}$ and $\mathcal{J}_{ji}^{\mathrm{read}}$ are defined as follows:
\begin{itemize}
\item For each nonconstant gate $i$, we introduce a set of items $\mathcal J_i^{\mathsf{ctl}}$, called its \emph{controller}, as follows:
\[
 \mathcal J_i^{\mathsf{ctl}}=
 \begin{cases}
  \{s_i^{(1)},s_i^{(2)},s_i^{\mathsf{Guard}}\},&\text{if gate }i\text{ is an }H\text{-gate},\\[2pt]
  \{s_i^{(1)},s_i^{(2)},\bar{s}_i^{(2)},s_i^{\mathsf{Guard}}\},&\text{if gate }i\text{ is a }G_{\mathrm{GSP}}\text{-gate}.
 \end{cases}
\]
\item For each readout $(j,i)$, we introduce the item set $\mathcal J_{ji}^{\mathsf{read}}=\{r_{ji}^{(1)},r_{ji}^{(2)},r_{ji}^{(3)}\}$. These item sets are pairwise disjoint.
\end{itemize}

Using positive rational base scales $V_i^{\mathsf{ctl}}$ and $V_{ji}^{\mathsf{read}}$, specified below in \eqref{eq:gsp-scales-full} and \eqref{eq:gsp-weight-definition}, respectively, assign the owners and values as in \cref{tab:gsp-items}.
For each readout $(j,i)$, let $U_{ji}$ denote its \emph{input bidder}, which represents $x_j$ and owns the item $r_{ji}^{(2)}$ in \cref{tab:gsp-items}. For each nonconstant variable $x_j$, enumerate the readouts $(j,i_1),\ldots,(j,i_d)$, where $0\leq d\leq2$, and set $U_{j,i_h}=A_j^{(h)}$ for $h\in[d]$. For each readout $(j,i)$ with constant $x_j$, set $U_{ji}=Q_j$. Each $A_j^{(1)},A_j^{(2)},Q_j$ therefore serves as $U_{ji}$ for at most one readout $(j,i)$.
\begin{table}[H]
\centering
\renewcommand{\arraystretch}{1.2}
\begin{tabular}{lll}
\toprule
Item & Owner & Value \\
\midrule
$s_i^{(1)}$ & $A_i^{(1)}$ & $V_i^{\mathsf{ctl}}$ \\
$s_i^{(2)}$ & $A_i^{(2)}$ & $V_i^{\mathsf{ctl}}$ \\
$s_i^{\mathsf{Guard}}$ & $\mathsf{Guard}$ & $V_i^{\mathsf{ctl}}/4$ \\
$\bar{s}_i^{(2)}$ & $A_i^{(2)}$ & $V_i^{\mathsf{ctl}}/2$ \\
\midrule
$r_{ji}^{(1)}$ &
$\begin{cases}
 A_i^{(1)},&\text{if gate }i\text{ is an }H\text{-gate},\\
 A_i^{(2)},&\text{if gate }i\text{ is a }G_{\mathrm{GSP}}\text{-gate}
\end{cases}$ & $4V_{ji}^{\mathsf{read}}$ \\
$r_{ji}^{(2)}$ & $U_{ji}$ & $V_{ji}^{\mathsf{read}}$ \\
$r_{ji}^{(3)}$ & $\mathsf{Guard}$ & $V_{ji}^{\mathsf{read}}/4$ \\
\bottomrule
\end{tabular}
\caption{Items in the GSP construction.}
\label{tab:gsp-items}
\end{table}

\paragraph{Parameters and Budgets.}
Set $K=128$. For each gate $i$, define its encoding width $\delta_i$; for each nonconstant gate, also define its controller scale $V_i^{\mathsf{ctl}}$:
\begin{equation}\label{eq:gsp-scales-full}
 \delta_i=\frac{1}{4K^{t_i}},
 \qquad V_i^{\mathsf{ctl}}=K^{5i}.
\end{equation}
In particular, $\delta_i=1/4$ on the original coordinates and on every $G_{\mathrm{GSP}}$-gate.

For each readout $(j,i)$, set
\begin{equation}\label{eq:gsp-weight-definition}
 V_{ji}^{\mathsf{read}}=
 \begin{cases}
 \frac{\delta_i}{\delta_j}V_i^{\mathsf{ctl}},&\text{if gate }i\text{ is an }H\text{-gate},\\[2pt]
 \frac{2}{5\delta_j}V_i^{\mathsf{ctl}},&\text{if gate }i\text{ is a }G_{\mathrm{GSP}}\text{-gate}.
 \end{cases}
\end{equation}

Assign each bidder its budget in \cref{tab:gsp-parameters}. In the table, each sum over $i$ ranges over readouts $(j,i)$ satisfying the displayed condition; an empty sum is zero.
\begin{table}[H]
\centering
\renewcommand{\arraystretch}{1.35}
\begin{tabular}{lll}
\toprule
Bidder $\ell$ & Gate $j$ & Budget $M_\ell$ \\
\midrule
$A_j^{(1)}$ & $H$-gate & $(3/4-\delta_j)V_j^{\mathsf{ctl}}+\sum_{k\in I_j}V_{kj}^{\mathsf{read}}+\frac14\sum_{i:\,U_{ji}=A_j^{(1)}}V_{ji}^{\mathsf{read}}$ \\
$A_j^{(2)}$ & $H$-gate & $V_j^{\mathsf{ctl}}/2+\frac14\sum_{i:\,U_{ji}=A_j^{(2)}}V_{ji}^{\mathsf{read}}$ \\
\midrule
$A_j^{(1)}$ & $G_{\mathrm{GSP}}$-gate & $3V_j^{\mathsf{ctl}}/5+\frac14\sum_{i:\,U_{ji}=A_j^{(1)}}V_{ji}^{\mathsf{read}}$ \\
$A_j^{(2)}$ & $G_{\mathrm{GSP}}$-gate & $7V_j^{\mathsf{ctl}}/10+V_{kj}^{\mathsf{read}}+\frac14\sum_{i:\,U_{ji}=A_j^{(2)}}V_{ji}^{\mathsf{read}}$ \\
\midrule
$Q_j$ & Constant gate & $2\sum_{i:\,U_{ji}=Q_j}V_{ji}^{\mathsf{read}}$ \\
\midrule
$\mathsf{Guard}$ & --- & $1+\sum_{s\in\mathcal J_{\mathsf{Guard}}}v_s$ \\
\bottomrule
\end{tabular}
\caption{Budgets in the GSP construction. For a $G_{\mathrm{GSP}}$-gate, $k$ is its input index.}
\label{tab:gsp-parameters}
\end{table}

This defines an auction instance $\mathcal I_{\Gc}$. All values and budgets are positive, since $0<\delta_i\leq1/4$. There are $O(N_0)$ bidders and items. The powers $K^{5i}$ have $O(N)$ bits, so all parameters have polynomial bit length and are computable in polynomial time.

\subsection{Scale Separation}\label{subsec:gsp-scales}
\begin{lemma}[Multiplier bounds]\label{lem:gsp-multipliers}
Every budget-constrained autobidding equilibrium $(\bm{\alpha},\pi)$ of $\mathcal I_{\Gc}$ satisfies $\alpha_{\mathsf{Guard}}=\alpha_{Q_i}=1$ for every constant input bidder $Q_i$, and $\alpha_{\ell}>1/4$ for every other bidder $\ell$.
\end{lemma}
\begin{proof}
Fix an equilibrium $(\bm{\alpha},\pi)$ of $\mathcal I_{\Gc}$. First, $\alpha_{\mathsf{Guard}}=1$.
For every ranking $\sigma=(s_1,\ldots,s_m)$ in the support of $\pi$ and every position $k\in[m]$, \eqref{eq:gsp-ranking} and \eqref{eq:gsp-item-price} give the price bound $0\leq c_k(\bm{\alpha},\sigma)\leq b_{s_k}$, where $b_{s_k}=\alpha_\ell v_{s_k}$ is the bid on the item $s_k$ in position $k$ and $\ell$ is its owner. Substituting $b_s=\alpha_{\ell} v_s$ for $s\in\mathcal J_{\ell}$ into \eqref{eq:gsp-total-payment} gives, for every bidder $\ell$,
\begin{equation}\label{eq:gsp-unconditional-pay}
 P_{\ell}(\bm{\alpha},\pi)\leq\alpha_{\ell}\sum_{s\in\mathcal J_{\ell}}v_s.
\end{equation}
For the guard, the budget in \cref{tab:gsp-parameters} gives
$P_{\mathsf{Guard}}(\bm{\alpha},\pi)
 \leq\sum_{s\in\mathcal J_{\mathsf{Guard}}}v_s
 =M_{\mathsf{Guard}}-1<M_{\mathsf{Guard}}$,
so \eqref{eq:gsp-unused-budget} forces $\alpha_{\mathsf{Guard}}=1$ (recall that $M_{\mathsf{Guard}}$ is the budget of $\mathsf{Guard}$).

For each constant input bidder $Q_j$, let $(j,i)$ be the unique readout with first coordinate $j$, so $U_{ji}=Q_j$. Then $\mathcal J_{Q_j}=\{r_{ji}^{(2)}\}$ and $v_{r_{ji}^{(2)}}=V_{ji}^{\mathsf{read}}<M_{Q_j}=2V_{ji}^{\mathsf{read}}$. Equation~\eqref{eq:gsp-unconditional-pay} therefore gives $P_{Q_j}(\bm{\alpha},\pi)<M_{Q_j}$, so \eqref{eq:gsp-unused-budget} yields $\alpha_{Q_j}=1$.

Next, fix a nonconstant gate $j$ and $\ell\in\{A_j^{(1)},A_j^{(2)}\}$. If $(j,i)$ is a readout with $U_{ji}=\ell$, then $\ell$ owns $r_{ji}^{(2)}$, of value $V_{ji}^{\mathsf{read}}$, and its budget in \cref{tab:gsp-parameters} contains the corresponding summand $V_{ji}^{\mathsf{read}}/4$. The contribution of readout $(j,i)$ to $M_\ell-\frac14\sum_{s\in\mathcal J_\ell}v_s$ is
\[
 \frac{V_{ji}^{\mathsf{read}}}{4}-\frac14V_{ji}^{\mathsf{read}}=0.
\]
There is at most one such gate $i$ for each bidder $\ell$. Substituting the item values from \cref{tab:gsp-items} and the budgets from \cref{tab:gsp-parameters} gives
\[
\begin{array}{c|c|c}
 \text{Gate type}&\ell&\bigl(M_\ell-\frac14\sum_{s\in\mathcal J_\ell}v_s\bigr)/V_j^{\mathsf{ctl}}\\ \hline
 H\text{-gate}&A_j^{(1)}&1/2-\delta_j\ \geq\ 1/4\\[2pt]
 H\text{-gate}&A_j^{(2)}&1/4\\[2pt]
 G_{\mathrm{GSP}}\text{-gate}&A_j^{(1)}&7/20\\[2pt]
 G_{\mathrm{GSP}}\text{-gate}&A_j^{(2)}&13/40
\end{array}
\]
If $\alpha_{\ell}\leq1/4$, \eqref{eq:gsp-unconditional-pay} would give
\[
 P_{\ell}(\bm{\alpha},\pi)
 \leq\alpha_{\ell}\sum_{s\in\mathcal J_{\ell}}v_s
 \leq\frac14\sum_{s\in\mathcal J_{\ell}}v_s<M_{\ell},
\]
where we used $M_\ell>\frac{1}{4}\sum_{s\in\mathcal J_{\ell}}v_s$ by the table above.
Now, \eqref{eq:gsp-unused-budget} would force $\alpha_{\ell}=1$, contradicting $\alpha_{\ell}\leq1/4$.
\end{proof}

\begin{lemma}[Scale separation]\label{lem:gsp-scales}
In every budget-constrained autobidding equilibrium $(\bm{\alpha},\pi)$ of $\mathcal I_{\Gc}$, the items in each controller $\mathcal J_i^{\mathsf{ctl}}$ and in each readout item set $\mathcal J_{ji}^{\mathsf{read}}$ occupy consecutive positions in every ranking $\sigma$ with $\pi(\sigma)>0$.
\end{lemma}
\begin{proof}
Fix an equilibrium $(\bm{\alpha},\pi)$ of $\mathcal I_{\Gc}$. We show that distinct item sets have disjoint bid ranges.
By \eqref{eq:gsp-scales-full} and \eqref{eq:gsp-weight-definition}, the base scales are
\[
 V_i^{\mathsf{ctl}}=K^{5i},\qquad
 V_{ji}^{\mathsf{read}}=
 \begin{cases}
 K^{5i+t_j-t_i},&\text{for an }H\text{-gate},\\
 \frac85K^{5i+1},&\text{for a }G_{\mathrm{GSP}}\text{-gate}.
 \end{cases}
\]
For each $H$-gate $x_i=H(x_j,x_k)$, we have $\{t_i,t_j,t_k\}=\{0,1,2\}$, so $t_j-t_i$ and $t_k-t_i$ are distinct elements of $\{-2,-1,1,2\}$. Hence its two readout exponents and controller exponent are distinct and lie in $[5i-2,5i+2]$. A $G_{\mathrm{GSP}}$-gate has exponents $5i$ and $5i+1$. These integer intervals are disjoint for different gates, so all exponents are distinct. The coefficients are $1$ or $8/5$; thus increasing the exponent increases the base scale by a factor at least $5K/8=80$.

By construction (\cref{tab:gsp-items}), the value of every item in a controller or readout item set with base scale $s$ is $s/4$, $s/2$, $s$, or $4s$, and hence lies between $s/4$ and $4s$. Combining these value bounds with the multiplier bounds in \cref{lem:gsp-multipliers} gives, for every item $r$ in the set,
\[
 \frac{s}{4}\leq v_r\leq4s\quad\text{and}\quad
 \frac{s}{16}<b_r\leq4s.
\] 
For any two distinct controller or readout item sets, write their base scales as $s_1,s_2$ in increasing exponent order. The choice of $K$ gives $s_2/s_1\geq80$, hence $s_2/16>4s_1$. Thus every bid on an item in the set with the larger exponent is strictly greater than every bid on an item in the other set, including guard bids. By \eqref{eq:gsp-ranking}, the items in each controller $\mathcal J_i^{\mathsf{ctl}}$ and in each readout item set $\mathcal J_{ji}^{\mathsf{read}}$ therefore occupy consecutive positions in every ranking $\sigma$ with $\pi(\sigma)>0$.
\end{proof}

\subsection{Correctness}\label{subsec:gsp-composition}

The following three lemmas determine the readout payments and the equations for $H$-gates and $G_{\mathrm{GSP}}$-gates in the full auction $\mathcal I_{\Gc}$ constructed in \cref{subsec:gsp-construction}. We then recover a circuit solution in the proof of \cref{thm:gsp-complete}. All expectations below are over the ranking distribution $\pi$ of the equilibrium under consideration.

For each nonconstant gate $i$ and bidder $\ell\in\{A_i^{(1)},A_i^{(2)}\}$, define its \emph{normalized controller payment} by
\[
 \widetilde P_\ell^{\mathsf{ctl}}
 =\frac{1}{V_i^{\mathsf{ctl}}}
 \mathbb E_{\sigma=(s_1,\ldots,s_m)\sim\pi}
 \left[\sum_{\substack{k\in[m]\\s_k\in\mathcal J_\ell\cap\mathcal J_i^{\mathsf{ctl}}}}
 c_k(\bm{\alpha},\sigma)\right].
\]

\begin{lemma}[Readout payments]\label{lem:gsp-readout}
In every equilibrium $(\bm{\alpha},\pi)$ of $\mathcal I_{\Gc}$, fix a readout $(j,i)$ and let $R=A_i^{(1)}$ if gate $i$ is an $H$-gate and $R=A_i^{(2)}$ otherwise, as in \cref{tab:gsp-items}. Write $P_{\ell}^{\mathsf{read}}$ for bidder $\ell$'s expected payment on its items in $\mathcal J_{ji}^{\mathsf{read}}$. Then
\[
 P_R^{\mathsf{read}}=V_{ji}^{\mathsf{read}}\alpha_{U_{ji}}\quad\text{and}\quad P_{U_{ji}}^{\mathsf{read}}=V_{ji}^{\mathsf{read}}/4.
\]
\end{lemma}
\begin{proof}
By \cref{lem:gsp-multipliers}, $\alpha_{\mathsf{Guard}}=1$ and $\alpha_R,\alpha_{U_{ji}}>1/4$, so the bids on $r_{ji}^{(1)},r_{ji}^{(2)},r_{ji}^{(3)}$ satisfy
\[
 4V_{ji}^{\mathsf{read}}\alpha_R> V_{ji}^{\mathsf{read}}\alpha_{U_{ji}}>\frac{V_{ji}^{\mathsf{read}}}4
 =\frac{V_{ji}^{\mathsf{read}}}4\alpha_{\mathsf{Guard}}.
\]
By \eqref{eq:gsp-ranking} and \cref{lem:gsp-scales}, these three items occupy consecutive positions in this order. Their owners $R,U_{ji},\mathsf{Guard}$ are distinct because \cref{lem:bounded-copy} gives distinct input and output indices. Thus \eqref{eq:gsp-item-price} prices $r_{ji}^{(1)}$ at the bid on $r_{ji}^{(2)}$ and prices $r_{ji}^{(2)}$ at the guard's bid on $r_{ji}^{(3)}$. This gives the claimed payments.
\end{proof}

Any item in $\mathcal J_i^{\mathsf{ctl}}$ whose price is determined by an item outside $\mathcal J_i^{\mathsf{ctl}}$ has price at most $V_i^{\mathsf{ctl}}/4$. Indeed, by \cref{lem:gsp-scales}, an outside item $t$ ranked below an item of $\mathcal J_i^{\mathsf{ctl}}$ lies below all its items. Hence \eqref{eq:gsp-ranking} gives $b_t\leq b_{s_i^{\mathsf{Guard}}}=V_i^{\mathsf{ctl}}/4$, and \eqref{eq:gsp-item-price} gives the claimed price bound, including price zero when no lower-ranked item has a different owner. An item ranked at or below $s_i^{\mathsf{Guard}}$ also has price at most $V_i^{\mathsf{ctl}}/4$, since its price is at most its own bid.

For every readout $(j,i)$ with nonconstant $x_j$, \cref{lem:gsp-readout} gives payment $V_{ji}^{\mathsf{read}}/4$ by $U_{ji}$, equal to the corresponding summand in its budget in \cref{tab:gsp-parameters}. Thus payments made as input bidders cancel against these budget summands. After subtracting the remaining readout payments, we obtain the controller budget bounds used in the next two proofs. By \eqref{eq:gsp-tight-budget}, each bound is an equality whenever the corresponding bidder's multiplier is less than one.

\begin{lemma}[$H$-gate equations]\label{lem:gsp-affine}
In every equilibrium $(\bm{\alpha},\pi)$ of $\mathcal I_{\Gc}$, each $H$-gate $i$ satisfies
\[
 \alpha_{A_i^{(1)}}=\alpha_{A_i^{(2)}}
 =\min\left\{1,1-\delta_i+\sum_{j\in I_i}\frac{\delta_i}{\delta_j}(1-\alpha_{U_{ji}})\right\}
 \in[1-\delta_i,1].
\]
\end{lemma}
\begin{proof}
Fix an equilibrium $(\bm{\alpha},\pi)$ and an $H$-gate $i$. By \cref{lem:gsp-readout}, $A_i^{(1)}$ pays $\sum_{j\in I_i}V_{ji}^{\mathsf{read}}\alpha_{U_{ji}}$ on the readout items introduced for gate $i$, while $A_i^{(2)}$ owns none of these items. After canceling the payments made as input bidders, subtract these readout payments from the budgets in \cref{tab:gsp-parameters} and divide by $V_i^{\mathsf{ctl}}$. Using $V_{ji}^{\mathsf{read}}/V_i^{\mathsf{ctl}}=\delta_i/\delta_j$ gives
\begin{equation}\label{eq:gsp-affine-controller-budgets}
\begin{aligned}
 \widetilde P_{A_i^{(1)}}^{\mathsf{ctl}}
 &\leq\frac34-\delta_i+\sum_{j\in I_i}\frac{\delta_i}{\delta_j}(1-\alpha_{U_{ji}}),\\
 \widetilde P_{A_i^{(2)}}^{\mathsf{ctl}}&\leq\frac12.
\end{aligned}
\end{equation}
Both right-hand sides are at least $1/2$, since $\delta_i\leq1/4$ and every input multiplier is at most one.

Suppose $\alpha_{A_i^{(1)}}\ne\alpha_{A_i^{(2)}}$, and let $\ell\in\{A_i^{(1)},A_i^{(2)}\}$ have the smaller multiplier. Its controller item follows the other bidder's item in every supported ranking. Only the guard or an item below the controller can set its price, so $\widetilde P_\ell^{\mathsf{ctl}}\leq1/4$. Since $\alpha_\ell<1$, equality in its budget bound would give $\widetilde P_\ell^{\mathsf{ctl}}\geq1/2$, a contradiction. Thus $\alpha_{A_i^{(1)}}=\alpha_{A_i^{(2)}}$.

The tied items $s_i^{(1)},s_i^{(2)}$ lie above $s_i^{\mathsf{Guard}}$. In either order, their prices divided by $V_i^{\mathsf{ctl}}$ are $\alpha_{A_i^{(1)}}$ and $1/4$, so
\[
 \widetilde P_{A_i^{(1)}}^{\mathsf{ctl}}+\widetilde P_{A_i^{(2)}}^{\mathsf{ctl}}
 =\alpha_{A_i^{(1)}}+\frac14.
\]
Adding the two inequalities in \eqref{eq:gsp-affine-controller-budgets} and subtracting $1/4$ gives
\[
 \alpha_{A_i^{(1)}}
 \leq1-\delta_i+\sum_{j\in I_i}\frac{\delta_i}{\delta_j}(1-\alpha_{U_{ji}}),
\]
with equality whenever $\alpha_{A_i^{(1)}}<1$. Together with $\alpha_{A_i^{(1)}}\leq1$, this gives the asserted minimum formula. Its right-hand side is at least $1-\delta_i$, since every input multiplier is at most one.
\end{proof}

For $t>0$, define $\Phi(t)$ by
\begin{equation}\label{eq:gsp-B}
 \Phi(t)=2t+\frac{3}{10t}-\frac65.
\end{equation}

\begin{lemma}[$G_{\mathrm{GSP}}$-gate equations]\label{lem:gsp-nonlinear}
In every equilibrium $(\bm{\alpha},\pi)$ of $\mathcal I_{\Gc}$, each $G_{\mathrm{GSP}}$-gate $x_i=G_{\mathrm{GSP}}(x_j)$ satisfies $\alpha_{A_i^{(1)}}=\alpha_{A_i^{(2)}}\in[3/4,1]$ and
\[
 \Phi(\alpha_{A_i^{(1)}})
 =\frac7{10}+\frac{2(1-\alpha_{U_{ji}})}{5\delta_j}.
\]
\end{lemma}
\begin{proof}
Fix an equilibrium $(\bm{\alpha},\pi)$ and a $G_{\mathrm{GSP}}$-gate $x_i=G_{\mathrm{GSP}}(x_j)$. Since $t_j=1$, gate $j$ is an $H$-gate or a constant gate. By \cref{lem:gsp-affine,lem:gsp-multipliers}, its input bidder satisfies $1-\delta_j\leq\alpha_{U_{ji}}\leq1$.

By \cref{lem:gsp-readout}, $A_i^{(2)}$ pays $V_{ji}^{\mathsf{read}}\alpha_{U_{ji}}$ on the readout items introduced for gate $i$, while $A_i^{(1)}$ owns none of these items. After canceling the payments made as input bidders, subtract this payment from the budgets in \cref{tab:gsp-parameters} and divide by $V_i^{\mathsf{ctl}}$. Using $V_{ji}^{\mathsf{read}}/V_i^{\mathsf{ctl}}=2/(5\delta_j)$ gives
\begin{equation}\label{eq:gsp-nonlinear-controller-budgets}
\begin{aligned}
 \widetilde P_{A_i^{(1)}}^{\mathsf{ctl}}&\leq\frac35,\\
 \widetilde P_{A_i^{(2)}}^{\mathsf{ctl}}&\leq\frac7{10}+\frac{2(1-\alpha_{U_{ji}})}{5\delta_j}.
\end{aligned}
\end{equation}
Both right-hand sides are at least $3/5$, and the second is at most $11/10$.

Suppose the multipliers differ, and let $\ell$ be the bidder with the smaller multiplier. Its normalized controller payment is at most $1/2$: if $\ell=A_i^{(1)}$, its single controller item can be priced only by $\bar{s}_i^{(2)}$, the guard, or an outside item, giving price at most $V_i^{\mathsf{ctl}}/2$; if $\ell=A_i^{(2)}$, both controller items lie below $s_i^{(1)}$ and cannot price one another, so each has price at most $V_i^{\mathsf{ctl}}/4$. Since $\alpha_\ell<1$, its bound in \eqref{eq:gsp-nonlinear-controller-budgets} must be tight, requiring a normalized payment of at least $3/5$, a contradiction.

Write $\alpha=\alpha_{A_i^{(1)}}=\alpha_{A_i^{(2)}}$. By \eqref{eq:gsp-item-price}, $\widetilde P_{A_i^{(1)}}^{\mathsf{ctl}}\leq\alpha$. Thus $\alpha<3/5$ would contradict tightness of the first bound in \eqref{eq:gsp-nonlinear-controller-budgets}, so $\alpha\geq3/5$. In particular, $\alpha/2>1/4$, and the tied items $s_i^{(1)},s_i^{(2)}$ precede $\bar{s}_i^{(2)}$ and $s_i^{\mathsf{Guard}}$ within the controller.

For each ranking in the support of $\pi$, \cref{lem:gsp-scales} and \eqref{eq:gsp-item-price} give the controller payments of $A_i^{(1)},A_i^{(2)}$, divided by $V_i^{\mathsf{ctl}}$, as $(\alpha,1/2)$ when $s_i^{(1)}$ precedes $s_i^{(2)}$, and $(\alpha/2,\alpha+1/4)$ in the opposite order. In either order, multiplying the first payment by $2-1/(2\alpha)$ and adding the second gives $2\alpha$. Taking expectations over $\pi$ therefore yields
\begin{equation}\label{eq:gsp-nonlin-payments-full}
 \widetilde P_{A_i^{(2)}}^{\mathsf{ctl}}
 +\left(2-\frac1{2\alpha}\right)\widetilde P_{A_i^{(1)}}^{\mathsf{ctl}}
 =2\alpha.
\end{equation}

If either controller budget bound in \eqref{eq:gsp-nonlinear-controller-budgets} were strict, \eqref{eq:gsp-unused-budget} would force $\alpha=1$. Equation~\eqref{eq:gsp-nonlin-payments-full} and the upper bound $11/10$ on the second right-hand side would then give
\[
 2=\widetilde P_{A_i^{(2)}}^{\mathsf{ctl}}
   +\frac32\widetilde P_{A_i^{(1)}}^{\mathsf{ctl}}
 <\frac{11}{10}+\frac32\cdot\frac35=2,
\]
a contradiction. Hence both controller budget bounds are equalities. Substituting them into \eqref{eq:gsp-nonlin-payments-full} and using \eqref{eq:gsp-B} gives
\[
 \Phi(\alpha)
 =2\alpha-\left(2-\frac1{2\alpha}\right)\frac35
 =\frac7{10}+\frac{2(1-\alpha_{U_{ji}})}{5\delta_j}.
\]
Finally,
\[
 \frac{(4\alpha-3)(5\alpha-1)}{10\alpha}
 =\Phi(\alpha)-\frac7{10}
 =\frac{2(1-\alpha_{U_{ji}})}{5\delta_j}\geq0.
\]
Since $\alpha\geq3/5$, the factors $5\alpha-1$ and $10\alpha$ are positive, so $\alpha\geq3/4$.
\end{proof}

\begin{proof}[Proof of \cref{thm:gsp-complete}]
Given an $N_0$-gate $\{1,H,G_{\mathrm{GSP}}\}$-generalized circuit $\Gc_0$, let $\Gc$ and $\mathcal I_{\Gc}$ be the circuit and auction constructed in \cref{subsec:gsp-construction}.

Let $(\bm{\alpha},\pi)$ be any equilibrium of $\mathcal I_{\Gc}$, whose existence follows from~\cite{HLN26}. By \cref{lem:gsp-multipliers,lem:gsp-affine,lem:gsp-nonlinear}, every nonconstant gate satisfies $\alpha_{A_i^{(1)}}=\alpha_{A_i^{(2)}}\in[1-\delta_i,1]$, and the guard and constant input bidders have multiplier one. Define
\begin{equation}\label{eq:gsp-recovery}
 x_i=
 \begin{cases}
 (\alpha_{A_i^{(1)}}-(1-\delta_i))/\delta_i,&\text{if gate }i\text{ is nonconstant},\\
 \alpha_{\mathsf{Guard}},&\text{if gate }i\text{ is constant}.
 \end{cases}
\end{equation}
Then $\bm{x}\in[0,1]^N$, constant gates are satisfied, and every input bidder satisfies $\alpha_{U_{ji}}=1-\delta_j+\delta_jx_j$. For an $H$-gate $x_i=H(x_j,x_k)$, substituting this identity into \cref{lem:gsp-affine} gives
\[
 \alpha_{A_i^{(1)}}=\min\{1,1-\delta_i+\delta_i(2-x_j-x_k)\},
\]
and hence $x_i=H(x_j,x_k)$.

For every $G_{\mathrm{GSP}}$-gate $i$ with input $j$, \cref{lem:gsp-nonlinear} and $\alpha_{U_{ji}}=1-\delta_j+\delta_jx_j$ give

\[
 \Phi(\alpha_{A_i^{(1)}})=\frac7{10}+\frac25(1-x_j)=\frac{11}{10}-\frac25x_j.
\]

Since $\delta_i=1/4$, \eqref{eq:gsp-recovery} gives $\alpha_{A_i^{(1)}}=(3+x_i)/4$. For every $x\in[0,1]$, \eqref{eq:gsp-B} and \eqref{eq:gsp-ginv} give
\[
 \Phi\left(\frac{3+x}{4}\right)
 =\frac3{10}+\frac{x}{2}+\frac{6}{5(3+x)}
 =\frac{11}{10}-\frac25G_{\mathrm{GSP}}^{-1}(x).
\]
Applying this identity with $x=x_i$ and using $\Phi(\alpha_{A_i^{(1)}})=11/10-2x_j/5$ yields $G_{\mathrm{GSP}}^{-1}(x_i)=x_j$, hence $x_i=G_{\mathrm{GSP}}(x_j)$, including at the endpoints.

Thus every decoded vector $\bm{x}$ belongs to $\Sol(\Gc)$. The lemmas use the total payments in $\mathcal I_{\Gc}$ and establish all multiplier intervals before decoding any gate. They therefore apply to directed cycles; self-inputs of $\Gc_0$ are removed by the copy preprocessing.

By \cref{lem:bounded-copy}, $(x_1,\ldots,x_{N_0})\in\Sol(\Gc_0)$. For each $i\in[N_0]$, the recovery formula is $x_i=4\alpha_{A_i^{(1)}}-3$ if gate $i$ is nonconstant and $x_i=\alpha_{\mathsf{Guard}}$ if it is constant, since $t_i=0$ gives $\delta_i=1/4$.

The construction has $O(N_0)$ bidders and items and is computable in polynomial time; the size, positivity, and bit-length bounds are proved in \cref{subsec:gsp-construction}. Since finding a solution to a $\{1,H,G_{\mathrm{GSP}}\}$-generalized circuit is $\FIXP$-complete by \cref{lem:gsp-gate}, the recovery above gives an $\SL$-reduction to the multiplier-only problem. Together with membership from \cref{subsec:gsp-model}, this proves the theorem.

\end{proof}
\section{Throttling in Second-Price Auctions}\label{sec:throttling}
We prove \cref{thm:intro-throttling} for the model of Chen, Kroer, and Kumar~\cite{CKK21}.

\subsection{Model}\label{subsec:throttling-model}
An instance specifies $n$ bidders, $m$ items, nonnegative rational bids $\bm{b}=(b_{\ell,s})_{\ell\in[n],s\in[m]}$, and positive rational budgets $\bm{M}=(M_\ell)_{\ell\in[n]}$. Bidder $\ell$ participates in the auctions for all $m$ items with probability $\theta_\ell\in[0,1]$, and otherwise participates in none; these participation decisions are independent across bidders. Each of the $m$ items is offered once in a second-price auction among participating bidders; a sole participant pays zero. Ties are resolved by a fixed priority order. The bids and budgets are fixed inputs; the participation probabilities are the equilibrium variables.

Let $P_\ell(\bm{\theta})$ be bidder $\ell$'s expected total payment over all $m$ items. For bidders $\ell,\ell'\in[n]$ and an item $s\in[m]$, write $\ell\succ_s\ell'$ when $\ell$ precedes $\ell'$ in decreasing bid order on item $s$, including the fixed tie-breaking priority. Write $\bm{\theta}_{-\ell}=(\theta_{\ell'})_{\ell'\in[n]\setminus\{\ell\}}$. With all bidder indices ranging over $[n]$, define
\begin{equation}\label{eq:throttle-conditional}
 C_\ell(\bm{\theta}_{-\ell})=
 \sum_{s=1}^m\ \sum_{\ell':\,\ell\succ_s\ell'}
 b_{\ell',s}\theta_{\ell'}
 \prod_{\substack{\ell''\ne\ell\\\ell''\succ_s\ell'}}(1-\theta_{\ell''}).
\end{equation}
Thus $P_\ell(\bm{\theta})=\theta_\ell C_\ell(\bm{\theta}_{-\ell})$. We call $C_\ell(\bm{\theta}_{-\ell})$ the \emph{conditional payment}: bidder $\ell$'s expected total payment when $\theta_\ell$ is set to $1$ and $\bm{\theta}_{-\ell}$ is fixed.

\begin{definition}[Second-price throttling equilibrium; {\cite[Definition~1]{CKK21}}]\label{def:throttling}
An \emph{exact second-price throttling equilibrium} is a vector $\bm{\theta}\in[0,1]^n$ satisfying, for every $\ell\in[n]$,
\begin{align}
 P_\ell(\bm{\theta})&\leq M_\ell,\label{eq:throttle-budget}\\
 P_\ell(\bm{\theta})<M_\ell&\ \Longrightarrow\ \theta_\ell=1.\label{eq:throttle-cap}
\end{align}
\end{definition}
The two equilibrium conditions are equivalent to
\begin{equation}\label{eq:throttle-response}
 \theta_\ell=\min\left\{1,\frac{M_\ell}{C_\ell(\bm{\theta}_{-\ell})}\right\}.
\end{equation}
The right-hand side is defined to be $1$ when $C_\ell(\bm{\theta}_{-\ell})=0$.
If $C_\ell(\bm{\theta}_{-\ell})\leq M_\ell$, a choice $\theta_\ell<1$ would give $P_\ell=\theta_\ell C_\ell<M_\ell$, contradicting \eqref{eq:throttle-cap}; hence $\theta_\ell=1$. If $C_\ell(\bm{\theta}_{-\ell})>M_\ell$, budget feasibility requires $\theta_\ell<1$, and \eqref{eq:throttle-cap} then gives $\theta_\ell C_\ell=M_\ell$. Conversely, these choices satisfy \eqref{eq:throttle-budget} and \eqref{eq:throttle-cap} in their respective cases.

\begin{theorem}[Second-price throttling]\label{thm:throttling-complete}
Given binary-encoded rational bids $\bm{b}\in\mathbb{Q}_{\geq0}^{n\times m}$ and budgets $\bm{M}\in\mathbb{Q}_{>0}^n$, the problem of finding an exact second-price throttling equilibrium $\bm{\theta}\in[0,1]^n$ is $\FIXP$-complete. Hardness holds even when each item has at most three positive bids and these bids are pairwise distinct.
\end{theorem}

Equilibrium existence and $\FIXP$-membership follow from~\cite[Theorem~8 and the following remark]{CKK21}. We prove hardness below.

\begin{remark}
In the stochastic formulation of Chen, Kroer, and Kumar~\cite[Section~2]{CKK21}, an item type $s\in[m]$ is drawn with probability $d_s$, and bidder $\ell$ bids $\tilde b_{\ell,s}$ on that type. Setting $b_{\ell,s}=d_s\tilde b_{\ell,s}$ gives the same expected payments as offering every item once with bids $b_{\ell,s}$: for each set of participating bidders, scaling all bids on item $s$ by $d_s$ scales its second price by $d_s$. Thus the model above uses their deterministic multi-item formulation.

\end{remark}

\subsection{Gate Construction}\label{subsec:throttling-construction}
By \cref{thm:universal,lem:auction-affine}, finding a solution of a $\{1,H,R_1\}$-generalized circuit is $\FIXP$-complete. Apply \cref{lem:bounded-copy} so that, for each $j$, there are at most two readouts $(j,i)$ if $x_j$ is nonconstant and at most one if it is constant, and each gate's input and output indices are pairwise distinct. Fix the resulting $N$-gate circuit $\Gc$.

Introduce one bidder $\mathsf{Guard}$, shared by all gates, with budget one. On every item on which it bids positively, its bid will be the lowest positive bid.

Every invocation of an item construction below uses fresh items. All unspecified bids are zero.
Auxiliary bidders are used only within the gate that introduces them. Only bidders representing circuit variables and $\mathsf{Guard}$ may be shared between gates.

\subsubsection{Basic Items}\label{subsubsec:throttling-elementary-items}
The following two item constructions are used in both the $H$-gates and $R_1$-gates defined in \cref{subsubsec:throttling-affine-gates}.
For distinct bidders $U,V$ and a rational $w>0$, let $\mathsf{Pos}(U,V;w)$ denote the single item in \cref{tab:throttle-basic-bids}.
For distinct bidders $U,V$, neither equal to $\mathsf{Guard}$, and rationals $k>d>0$, let $\mathsf{Neg}^{(1)}(U,V;k,d)$ and $\mathsf{Neg}^{(2)}(U,V;k,d)$ denote the two items in \cref{tab:throttle-basic-bids}, and write $\mathsf{Neg}(U,V;k,d)$ for their pair.
\begin{table}[H]
\centering
\renewcommand{\arraystretch}{1.2}
\begin{tabular}{lc@{\qquad}lcc}
\toprule
\multicolumn{2}{c}{$\mathsf{Pos}(U,V;w)$} & \multicolumn{3}{c}{$\mathsf{Neg}(U,V;k,d)$}\\
\cmidrule(r){1-2}\cmidrule(l){3-5}
Bidder & $\mathsf{Pos}$ & Bidder & $\mathsf{Neg}^{(1)}$ & $\mathsf{Neg}^{(2)}$\\
\midrule
$U$ & $w$ & $U$ & $2k$ & $k-d$\\
$V$ & $2w$ & $V$ & $k$ & $2k$\\
& & $\mathsf{Guard}$ & $(k+d)/2$ & $(k-d)/2$\\
\bottomrule
\end{tabular}
\caption{Fixed bids in $\mathsf{Pos}(U,V;w)$ and $\mathsf{Neg}(U,V;k,d)$, where $w>0$ and $k>d>0$. All unlisted bids are zero.}
\label{tab:throttle-basic-bids}
\end{table}

\subsubsection{$H$-Gates and $R_1$-Gates}\label{subsubsec:throttling-affine-gates}
For every gate $j$, introduce a bidder $X_j$, with the intended encoding $\theta_{X_j}=(1+x_j)/2$. A constant gate $x_j=1$ has only this bidder, with budget $M_{X_j}=2$. By the preprocessing in \cref{lem:bounded-copy}, there is at most one readout $(j,i)$, for which the input items are constructed as for any other variable.

For every nonconstant gate $i$, also introduce a bidder $Y_i$. For an $H$-gate $x_i=H(x_j,x_k)$, introduce one more bidder $Z_i$. The items and budgets are specified in \cref{tab:throttle-affine-summary}. For each readout $(j,i)$, the input bidder $X_j$ retains the budget assigned by gate $j$.
\begin{table}[H]
\centering
\small
\renewcommand{\arraystretch}{1.3}
\begin{tabular}{lll}
\toprule
Gate & Item constructions & Budgets \\
\midrule
$x_i=H(x_j,x_k)$
& $\mathsf{Neg}(X_j,Z_i;9/8,1)$ & $M_{X_i}=3$ \\
& $\mathsf{Neg}(X_k,Z_i;9/8,1)$ & $M_{Y_i}=13$ \\
& $\mathsf{Pos}(Z_i,Y_i;39/2)$ & $M_{Z_i}=1/8$ \\
& $\mathsf{Neg}(Y_i,X_i;39/4,9)$ & \\
\midrule
$x_i=R_1(x_j)$
& $\mathsf{Pos}(X_j,Y_i;39/2)$ & $M_{X_i}=6$ \\
& $\mathsf{Neg}(Y_i,X_i;39/4,9)$ & $M_{Y_i}=13$ \\
\midrule
$x_i=1$ & None in this gate & $M_{X_i}=2$ \\
\bottomrule
\end{tabular}
\caption{Items and budgets for all three gate types. Each invocation uses fresh items. Items for readouts of a constant variable are introduced by the gates that read it.}
\label{tab:throttle-affine-summary}
\end{table}

For each nonconstant variable $x_j$, add
\begin{equation}\label{eq:throttle-added-items}
 2-\left|\{i\in[N]:(j,i)\text{ is a readout and }g_i\text{ is an }H\text{-gate}\}\right|
\end{equation}
fresh items $\mathsf{Pos}(\mathsf{Guard},X_j;9/8)$. This number is nonnegative by \cref{lem:bounded-copy}. By \cref{lem:throttle-negative} below, each readout $(j,i)$ with an $H$-gate $i$ and each added item contributes $9/8$ to $X_j$'s conditional payment when the guard participates, whereas each readout $(j,i)$ with an $R_1$-gate $i$ contributes zero. Thus their total is always $9/4$.

For each nonconstant gate $i$, the pair $\mathsf{Neg}(Y_i,X_i;39/4,9)$ contributes $39/4-9\theta_{Y_i}$ to $X_i$'s conditional payment. Including the readout and added-item contribution gives total conditional payment $9/4+(39/4-9\theta_{Y_i})=12-9\theta_{Y_i}$. This choice makes the response of an $H$-gate output equal to $1/(4-3\theta_{Y_i})$; doubling its budget gives the $R_1$-gate output. The budget $13$ of $Y_i$ gives the ratio $13/(39/4)=4/3$ in both constructions.

\begin{figure}[H]
\centering
\begin{tikzpicture}[
 font=\small,
 bidder/.style={circle,draw=blue!55!black,fill=blue!8,minimum size=9mm,inner sep=2pt,line width=.65pt},
 item/.style={rectangle,draw=black!65,fill=black!3,minimum width=16mm,minimum height=9mm,inner sep=3pt,line width=.65pt,align=center,font=\footnotesize},
 link/.style={draw=black!70,line width=.75pt}
]
\node[font=\small\bfseries] at (0,1) {(a) $x_i=H(x_j,x_k)$};
\node[bidder] (j) at (-2,0) {$X_j$};
\node[bidder] (k) at (2,0) {$X_k$};
\node[item] (jneg1) at (-3,-1.25) {$\mathsf{Neg}^{(1)}$\\$(9/8,1)$};
\node[item] (jneg2) at (-1,-1.25) {$\mathsf{Neg}^{(2)}$\\$(9/8,1)$};
\node[item] (kneg1) at (1,-1.25) {$\mathsf{Neg}^{(1)}$\\$(9/8,1)$};
\node[item] (kneg2) at (3,-1.25) {$\mathsf{Neg}^{(2)}$\\$(9/8,1)$};
\node[bidder] (z) at (0,-2.75) {$Z_i$};
\node[item] (pos) at (0,-4) {$\mathsf{Pos}$\\$(39/2)$};
\node[bidder] (y) at (0,-5.25) {$Y_i$};
\node[item] (neg1) at (-1,-6.5) {$\mathsf{Neg}^{(1)}$\\$(39/4,9)$};
\node[item] (neg2) at (1,-6.5) {$\mathsf{Neg}^{(2)}$\\$(39/4,9)$};
\node[bidder] (x) at (0,-7.75) {$X_i$};
\draw[link] (j.south west) -- (jneg1.north) (jneg1.south) -- (z.155);
\draw[link] (j.south east) -- (jneg2.north) (jneg2.south) -- (z.115);
\draw[link] (k.south west) -- (kneg1.north) (kneg1.south) -- (z.65);
\draw[link] (k.south east) -- (kneg2.north) (kneg2.south) -- (z.25);
\draw[link] (z.south) -- (pos.north) (pos.south) -- (y.north);
\draw[link] (y.south west) -- (neg1.north) (neg1.south) -- (x.north west);
\draw[link] (y.south east) -- (neg2.north) (neg2.south) -- (x.north east);
\node[anchor=west] at (.65,-2.75) {$M_{Z_i}=1/8$};
\node[anchor=west] at (.65,-5.25) {$M_{Y_i}=13$};
\node[anchor=west] at (.65,-7.75) {$M_{X_i}=3$};
\node[font=\small\bfseries] at (7.7,1) {(b) $x_i=R_1(x_j)$};
\node[bidder] (rj) at (7.7,0) {$X_j$};
\node[item] (rpos) at (7.7,-2.75) {$\mathsf{Pos}$\\$(39/2)$};
\node[bidder] (ry) at (7.7,-5.25) {$Y_i$};
\node[item] (rneg1) at (6.7,-6.5) {$\mathsf{Neg}^{(1)}$\\$(39/4,9)$};
\node[item] (rneg2) at (8.7,-6.5) {$\mathsf{Neg}^{(2)}$\\$(39/4,9)$};
\node[bidder] (rx) at (7.7,-7.75) {$X_i$};
\draw[link] (rj.south) -- (rpos.north) (rpos.south) -- (ry.north);
\draw[link] (ry.south west) -- (rneg1.north) (rneg1.south) -- (rx.north west);
\draw[link] (ry.south east) -- (rneg2.north) (rneg2.south) -- (rx.north east);
\node[anchor=west] at (8.35,-5.25) {$M_{Y_i}=13$};
\node[anchor=west] at (8.35,-7.75) {$M_{X_i}=6$};
\end{tikzpicture}
\caption{The fixed throttling gate constructions. Circles denote bidders, rectangles denote individual items, and edges denote positive bids. Item labels show the construction type and its parameters, with the upper bidder as $U$ and the lower bidder as $V$ in \cref{tab:throttle-basic-bids}. The guard and its incident edges are omitted; it bids on every $\mathsf{Neg}^{(1)}$ and $\mathsf{Neg}^{(2)}$ item as specified in that table. Additional items that make the conditional payment from input uses equal to $9/4$ are omitted. Input bidders are shared with their own gates.}
\label{fig:throttling-cycle-example}
\end{figure}
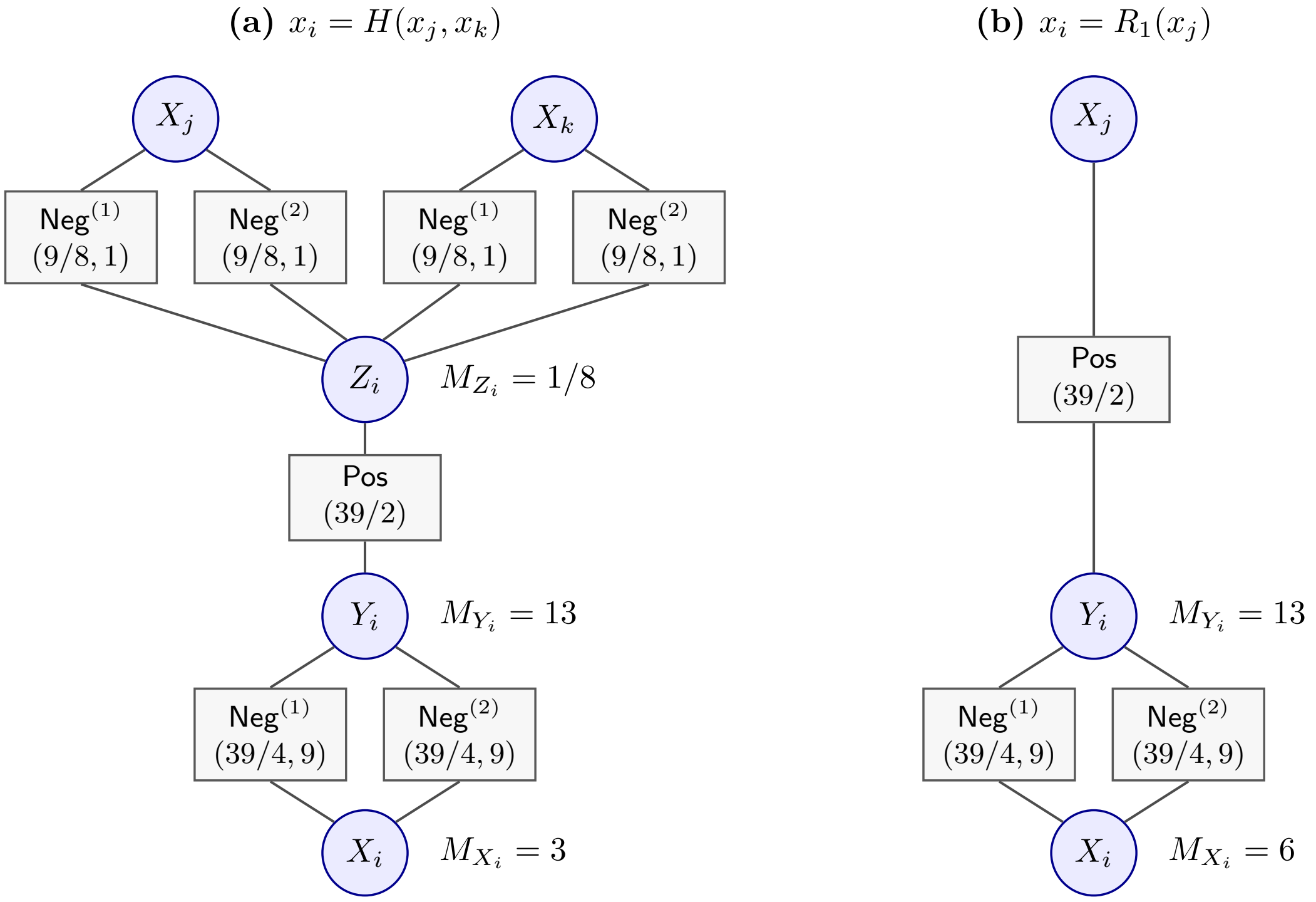

\Cref{fig:throttling-cycle-example} shows the two gate constructions after input copying. The basic items and their bids are specified in \cref{tab:throttle-basic-bids}.

Let $\mathcal B$ and $\mathcal J$ be the sets of bidders and items introduced above. Together with the specified bids and budgets, they define $\mathcal I_{\Gc}$. Write $\bm{\theta}=(\theta_\ell)_{\ell\in\mathcal B}$ for its participation vector. For each $H$-gate $x_i=H(x_j,x_k)$, define
\begin{equation}\label{eq:throttle-affine-expression}
 F_i=\frac52-\theta_{X_j}-\theta_{X_k}.
\end{equation}
Note that $F_i\geq1/2$ for arbitrary participation probabilities.

\subsection{Correctness}\label{subsec:throttling-correctness}

\subsubsection{Size and Construction Time}\label{subsubsec:throttling-size}
An $H$-gate uses three bidders and seven items, an $R_1$-gate uses two bidders and three items, and a constant gate uses one bidder. At most two additional items are introduced for each nonconstant gate. Including the shared guard, the auction therefore has at most $3N+1$ bidders and $9N$ items.

All bids and budgets belong to fixed finite sets of positive rationals, except for omitted bids, which are zero. The two $\mathsf{Neg}$ parameter pairs satisfy $9/8>1>0$ and $39/4>9>0$. Each $\mathsf{Pos}$ item has two distinct positive bids. In every $\mathsf{Neg}$ pair, the strict bid orders are
\[
 2k>k>\frac{k+d}{2}>0,
 \qquad 2k>k-d>\frac{k-d}{2}>0.
\]
Thus each item has at most three positive bids, all pairwise distinct. The construction is polynomial time and needs no ordering of the gates, including for circuits with directed cycles.

\subsubsection{Gate Properties}\label{subsubsec:throttling-gate-properties}
On every item where the guard bids positively, its bid is the lowest positive bid. If it wins, every other positive bidder is absent, so its price is zero. Hence $P_{\mathsf{Guard}}(\bm{\theta})=0<1=M_{\mathsf{Guard}}$, and \eqref{eq:throttle-cap} gives $\theta_{\mathsf{Guard}}=1$ at every equilibrium.

\begin{lemma}[Basic Item Payments]\label{lem:throttle-negative}
For the item $\mathsf{Pos}(U,V;w)$ defined in \cref{subsubsec:throttling-elementary-items}, let $P_U^{\mathsf{Pos}}$ and $P_V^{\mathsf{Pos}}$ denote the expected payments of $U$ and $V$, respectively, on its item. Then, for every $\theta_U,\theta_V\in[0,1]$,
\begin{equation}\label{eq:throttle-positive-payments}
 P_U^{\mathsf{Pos}}=0\quad\text{and}\quad
 P_V^{\mathsf{Pos}}=w\theta_U\theta_V.
\end{equation}
For the two items $\mathsf{Neg}(U,V;k,d)$ defined in \cref{subsubsec:throttling-elementary-items}, let $P_U^{\mathsf{Neg}}$ and $P_V^{\mathsf{Neg}}$ denote the expected total payments of $U$ and $V$, respectively, summed over its two items. If $\theta_{\mathsf{Guard}}=1$, then $P_U^{\mathsf{Neg}}=k\theta_U$ and $P_V^{\mathsf{Neg}}=\theta_V(k-d\theta_U)$ for every $\theta_U,\theta_V\in[0,1]$.
\end{lemma}
\begin{proof}
On $\mathsf{Pos}(U,V;w)$, bidder $V$ pays $w$ if both $U$ and $V$ participate; otherwise, all payments on this item are zero. Independence between bidders gives \eqref{eq:throttle-positive-payments}.

For $\mathsf{Neg}(U,V;k,d)$, assume $\theta_{\mathsf{Guard}}=1$. By \cref{tab:throttle-basic-bids}, the strict bid order is $U,V,\mathsf{Guard}$ on $\mathsf{Neg}^{(1)}(U,V;k,d)$ and $V,U,\mathsf{Guard}$ on $\mathsf{Neg}^{(2)}(U,V;k,d)$. With $U$ present, its combined payment is $k$ when $V$ is present and $(k+d)/2+(k-d)/2=k$ when $V$ is absent. With $V$ present, its combined payment is $k-d$ when $U$ is present and $(k+d)/2+(k-d)/2=k$ when $U$ is absent. Independence between bidders and linearity of expectation therefore give
\begin{equation}\label{eq:throttle-negative-payments}
\begin{aligned}
 P_U^{\mathsf{Neg}}
 &=\theta_U\bigl[\theta_V k+(1-\theta_V)k\bigr]=k\theta_U,\\
 P_V^{\mathsf{Neg}}
 &=\theta_V\bigl[\theta_U(k-d)+(1-\theta_U)k\bigr]
 =\theta_V(k-d\theta_U).
\end{aligned}
\end{equation}
Thus $U$'s payment is independent of $\theta_V$.

\end{proof}

\subsubsection{Connecting Gates}\label{subsec:throttling-composition}

\begin{lemma}\label{lem:throttle-composition-payments}
For every participation vector with $\theta_{\mathsf{Guard}}=1$, the nonconstant gates of $\mathcal I_{\Gc}$ have the total payments in \cref{tab:throttle-composition-payments}. At every equilibrium, each constant gate satisfies $\theta_{X_i}=1$.
\end{lemma}
\begin{table}[H]
\centering
\renewcommand{\arraystretch}{1.25}
\begin{tabular}{lll}
\toprule
Gate & Bidder & Total expected payment \\
\midrule
$H$-gate & $Z_i$ & $\theta_{Z_i}(F_i-1/4)$ \\
$H$-gate & $Y_i$ & $(39/4)\theta_{Y_i}(1+2\theta_{Z_i})$ \\
$R_1$-gate & $Y_i$ & $(39/4)\theta_{Y_i}(1+2\theta_{X_j})$ \\
Either & $X_i$ & $\theta_{X_i}(12-9\theta_{Y_i})$ \\
\bottomrule
\end{tabular}
\caption{Total payments in the full auction. For an $R_1$-gate, $j$ is its input index.}
\label{tab:throttle-composition-payments}
\end{table}
\begin{proof}
For each readout $(j,i)$ with an $H$-gate $i$, \cref{lem:throttle-negative} gives payment $(9/8)\theta_{X_j}$ by $X_j$ on $\mathsf{Neg}(X_j,Z_i;9/8,1)$, independently of $\theta_{Z_i}$. Each added item $\mathsf{Pos}(\mathsf{Guard},X_j;9/8)$ also contributes $(9/8)\theta_{X_j}$ because $\theta_{\mathsf{Guard}}=1$. For each readout $(j,i)$ with an $R_1$-gate $i$, bidder $X_j$ pays zero on $\mathsf{Pos}(X_j,Y_i;39/2)$. By \eqref{eq:throttle-added-items}, the number of readouts $(j,i)$ with an $H$-gate $i$ plus the number of added items for a nonconstant $X_j$ is exactly two. Adding its payment on $\mathsf{Neg}(Y_j,X_j;39/4,9)$ gives
\[
 P_{X_j}=\theta_{X_j}\left(\frac94+\frac{39}4-9\theta_{Y_j}\right)
 =\theta_{X_j}(12-9\theta_{Y_j}).
\]
For a constant gate $j$, there is at most one readout $(j,i)$, and bidder $X_j$ has no items from its own gate or added $\mathsf{Pos}(\mathsf{Guard},X_j;9/8)$ items. The readout payments just computed give $P_{X_j}(\bm{\theta})\leq(9/8)\theta_{X_j}\leq9/8<2=M_{X_j}$. Equation~\eqref{eq:throttle-cap} therefore gives $\theta_{X_j}=1$.

For an $H$-gate $i$ with inputs $j,k$, the pairs $\mathsf{Neg}(X_j,Z_i;9/8,1)$ and $\mathsf{Neg}(X_k,Z_i;9/8,1)$ give
\[
 P_{Z_i}=\theta_{Z_i}\left(\frac98-\theta_{X_j}+\frac98-\theta_{X_k}\right)
 =\theta_{Z_i}(F_i-1/4),
\]
since $Z_i$ pays zero on $\mathsf{Pos}(Z_i,Y_i;39/2)$.
For either nonconstant gate type, $Y_i$ pays $(39/4)\theta_{Y_i}$ on $\mathsf{Neg}(Y_i,X_i;39/4,9)$. Its additional payment is $(39/2)\theta_{Z_i}\theta_{Y_i}$ on $\mathsf{Pos}(Z_i,Y_i;39/2)$ for an $H$-gate, and $(39/2)\theta_{X_j}\theta_{Y_i}$ on $\mathsf{Pos}(X_j,Y_i;39/2)$ for an $R_1$-gate with input $j$. Each auxiliary bidder has zero bids outside the items of its own gate, so summing these contributions gives its total payment in \cref{tab:throttle-composition-payments}.
\end{proof}

\begin{lemma}[$H$-Gate Equations]\label{lem:throttle-template}
In every equilibrium of $\mathcal I_{\Gc}$, each $H$-gate $i$ satisfies
\[
 \theta_{Z_i}=\frac1{8F_i-2},\qquad
 \theta_{Y_i}=\min\left\{1,\frac43-\frac1{3F_i}\right\},\qquad
 \theta_{X_i}=\min\{1,F_i\}\in[1/2,1].
\]
\end{lemma}
\begin{proof}
Fix an equilibrium $\bm{\theta}$ and an $H$-gate $i$ with inputs $j,k$. By \eqref{eq:throttle-affine-expression}, $F_i=5/2-\theta_{X_j}-\theta_{X_k}\geq1/2$. Recall that $C_{Z_i}(\bm{\theta}_{-Z_i})$ is the conditional payment defined in \eqref{eq:throttle-conditional}. Thus \cref{lem:throttle-composition-payments} gives
\[
 C_{Z_i}(\bm{\theta}_{-Z_i})=F_i-\frac14\geq\frac14>M_{Z_i}=\frac18.
\]
Equation~\eqref{eq:throttle-response} gives $\theta_{Z_i}=1/(8F_i-2)$. Applying the same response equation with $M_{Y_i}=13$ and $C_{Y_i}(\bm{\theta}_{-Y_i})=(39/4)(1+2\theta_{Z_i})$, and substituting $\theta_{Z_i}=1/(8F_i-2)$ in the last equality, gives
\[
 \theta_{Y_i}
 =\min\left\{1,\frac{13}{(39/4)(1+2\theta_{Z_i})}\right\}
 =\min\left\{1,\frac43-\frac1{3F_i}\right\}.
\]
Since $12-9\theta_{Y_i}\geq3=M_{X_i}$, the output response is
\[
 \theta_{X_i}=\frac3{12-9\theta_{Y_i}}
 =\frac1{4-3\theta_{Y_i}}.
\]
If $1/2\leq F_i\leq1$, then $4/3-1/(3F_i)\in[2/3,1]$, so $\theta_{Y_i}=4/3-1/(3F_i)$ and $4-3\theta_{Y_i}=1/F_i$. Hence $\theta_{X_i}=F_i$. If $F_i\geq1$, then $4/3-1/(3F_i)\geq1$, so $\theta_{Y_i}=1$ and $\theta_{X_i}=1/(4-3)=1$. Both cases give $\theta_{X_i}=\min\{1,F_i\}\in[1/2,1]$, including at $F_i=1$.
\end{proof}

\begin{lemma}[$R_1$-Gate Equations]\label{lem:throttle-reciprocal}
In every equilibrium of $\mathcal I_{\Gc}$, each $R_1$-gate $x_i=R_1(x_j)$ satisfies
\[
 \theta_{Y_i}=\min\left\{1,\frac4{3(1+2\theta_{X_j})}\right\},\qquad
 \theta_{X_i}=\min\left\{1,\frac2{4-3\theta_{Y_i}}\right\}\in[3/4,1].
\]
If $\theta_{X_j}\geq1/2$, then $2\theta_{X_i}-1=1/(2\theta_{X_j})$.
\end{lemma}
\begin{proof}
For an $R_1$-gate $i$ with input $j$, \cref{lem:throttle-composition-payments} gives $C_{Y_i}(\bm{\theta}_{-Y_i})=(39/4)(1+2\theta_{X_j})$ and $C_{X_i}(\bm{\theta}_{-X_i})=12-9\theta_{Y_i}$. Substitution into \eqref{eq:throttle-response}, with $M_{Y_i}=13$ and $M_{X_i}=6$, gives the two stated response formulas. Since $\theta_{X_j}\leq1$, we have $\theta_{Y_i}\geq4/9$. Thus $1\leq4-3\theta_{Y_i}\leq8/3$ and $\theta_{X_i}=\min\{1,2/(4-3\theta_{Y_i})\}\geq3/4$.

If $\theta_{X_j}\in[1/2,1]$, then $4/[3(1+2\theta_{X_j})]\leq2/3<1$, so $\theta_{Y_i}=4/[3(1+2\theta_{X_j})]$. Substituting this value gives
\[
 \frac2{4-3\theta_{Y_i}}
 =\frac{1+2\theta_{X_j}}{4\theta_{X_j}}\in[3/4,1].
\]
Thus the cap in the response of $X_i$ also leaves this value unchanged, and
\[
 \theta_{X_i}=\frac{1+2\theta_{X_j}}{4\theta_{X_j}},\qquad
 2\theta_{X_i}-1=\frac1{2\theta_{X_j}}.
\]
These identities include $\theta_{X_j}=1/2$, where $\theta_{X_i}=1$.
\end{proof}

\begin{proof}[Proof of \cref{thm:throttling-complete}]
Given an $N_0$-gate $\{1,H,R_1\}$-generalized circuit $\Gc_0$, apply \cref{lem:bounded-copy} and let $\Gc$ be the resulting $N$-gate circuit, retaining the original variables at indices $[N_0]$. Let $\mathcal I_{\Gc}$ be the auction constructed in \cref{subsec:throttling-construction}.

Let $\bm{\theta}$ be any equilibrium of $\mathcal I_{\Gc}$. Such an equilibrium exists by~\cite[Theorem~8]{CKK21}, since all bids are nonnegative and all budgets are positive. The guard has $\theta_{\mathsf{Guard}}=1$. By \cref{lem:throttle-composition-payments,lem:throttle-template,lem:throttle-reciprocal}, we have $\theta_{X_i}\in[1/2,1]$ for every gate before decoding any input. For each $i\in[N]$, define
\begin{equation}\label{eq:throttle-recovery}
 x_i=2\theta_{X_i}-1\in[0,1].
\end{equation}
A constant bidder has $\theta_{X_i}=1$, so its recovered value is $1$. For an $H$-gate $x_i=H(x_j,x_k)$, substitution in \eqref{eq:throttle-affine-expression} gives
\[
 F_i=\frac{3-x_j-x_k}{2}.
\]
By \cref{lem:throttle-template},
\[
 x_i=2\min\{1,F_i\}-1=\min\{1,2-x_j-x_k\}=H(x_j,x_k).
\]
For an $R_1$-gate $i$ with input $j$, \cref{lem:throttle-reciprocal} and $\theta_{X_j}\geq1/2$ give
\[
 x_i=2\theta_{X_i}-1=\frac1{2\theta_{X_j}}=\frac1{1+x_j}=R_1(x_j).
\]
Thus $\bm{x}\in\Sol(\Gc)$. The payment identities in \cref{lem:throttle-composition-payments} hold in $\mathcal I_{\Gc}$ for every participation vector with $\theta_{\mathsf{Guard}}=1$, without assuming any gate equation. The decoding argument therefore also applies to directed cycles. By \cref{lem:bounded-copy}, restricting to indices $[N_0]$ gives $(x_i)_{i\in[N_0]}\in\Sol(\Gc_0)$.

By \cref{subsubsec:throttling-size}, the construction is polynomial time, uses positive rational budgets and nonnegative rational bids of constant bit length, and has at most three distinct positive bids per item. The source problem is $\FIXP$-complete by \cref{thm:universal,lem:auction-affine}. The recovery \eqref{eq:throttle-recovery} followed by projection is separable linear, proving hardness. Together with membership from \cref{subsec:throttling-model}, this proves the theorem.
\end{proof}

\section*{AI disclosure.} 
The basic framework in \cref{sec:framework} is originated from the author, which was later simplified with assistance from GPT-6 (particularly through the introduction of pseudo-gate for inversion).
The constructions of gadgets in \cref{sec:roi,sec:gsp,sec:throttling} are originated from GPT-6, where the author was aiming to find concrete applications of the framework of reciprocal gates. Later on, the author simplified the constructions with assistance from GPT-6.
The author verified the correctness and originality of all content, including references.

\clearpage
\appendix
\section{Verification of the Auction Gate Functions}\label{app:auction-gates}
This appendix gives the deferred proofs for the gate functions $G_{\mathrm{ROI}}$ and $G_{\mathrm{GSP}}$ used in \cref{sec:roi,sec:gsp}, respectively.

\begin{lemma}\label{lem:roi-gate}
Equation~\eqref{eq:roi-inverse} defines a continuous strictly decreasing bijection $G_{\mathrm{ROI}}:[0,1]\to[0,1]$. Finding a solution to a $\{1,H,G_{\mathrm{ROI}}\}$-generalized circuit is $\FIXP$-complete.
\end{lemma}
\begin{proof}
To justify defining $G_{\mathrm{ROI}}$ through its inverse, consider the function $x\mapsto2-\frac43x-R_{1/2}(x)$ on $[0,1]$. This function is continuous, equals $2-R_{1/2}(0)=1$ at $x=0$, equals $2-\frac43-R_{1/2}(1)=0$ at $x=1$, and has derivative
\[
 \frac{d}{dx}\left(2-\frac43x-R_{1/2}(x)\right)
 =-\frac43+\frac{1}{2(1+x/2)^2}\leq-\frac56<0.
\]
It is therefore a strictly decreasing bijection of $[0,1]$, whose inverse defines the continuous strictly decreasing bijection $G_{\mathrm{ROI}}$. Equation~\eqref{eq:roi-inverse} satisfies the hypotheses of \cref{cor:extraction} with $(a,b,c,\tau)=(2,-4/3,-1,1/2)$. Together with \cref{lem:auction-affine} and $1,H\in\A$, this proves the claimed $\FIXP$-completeness.
\end{proof}

\begin{lemma}\label{lem:gsp-gate}
Equation~\eqref{eq:gsp-ginv} defines a continuous strictly decreasing bijection $G_{\mathrm{GSP}}:[0,1]\to[0,1]$. Finding a solution to a $\{1,H,G_{\mathrm{GSP}}\}$-generalized circuit is $\FIXP$-complete.
\end{lemma}
\begin{proof}
The right-hand side of \eqref{eq:gsp-ginv} is continuous on $[0,1]$. At $x=0$ it equals $2-R_{1/3}(0)=1$, and at $x=1$ it equals $2-5/4-R_{1/3}(1)=0$. Its derivative is
\[
 \frac{d}{dx}\left(2-\frac54x-R_{1/3}(x)\right)
 =-\frac54+\frac{1}{3(1+x/3)^2}\leq-\frac{11}{12}<0.
\]
Its inverse $G_{\mathrm{GSP}}$ is therefore a continuous strictly decreasing bijection of $[0,1]$. The coefficients $(a,b,c,\tau)=(2,-5/4,-1,1/3)$ satisfy \cref{cor:extraction}, so \cref{lem:auction-affine} again gives the claimed $\FIXP$-completeness.
\end{proof}

\end{document}